\documentclass[12pt]{article}

\usepackage{fullpage}
\usepackage{amsmath}
\usepackage{graphicx}%
\usepackage{amsfonts}%
\usepackage{amssymb}
\usepackage{afterpage}
\usepackage{url}
\usepackage{fancyhdr}
\usepackage{isomath}
\usepackage{amsmath}
\usepackage{amsbsy}
\usepackage{amssymb}
\usepackage{amscd}
\usepackage{amsfonts}
\usepackage{graphicx}

\usepackage{verbatim}
\usepackage{euscript}
\usepackage{alltt}
\usepackage{stmaryrd}
\usepackage{float}
\usepackage{appendix}
\usepackage{caption}
\usepackage[english]{babel}
\usepackage{setspace} 
\usepackage{enumitem}
\usepackage{subcaption} 
\usepackage{layout}

\usepackage{tikz}

\usepackage{amsmath}
\usepackage{amsbsy}
\usepackage{amssymb}
\usepackage{amscd}
\usepackage{amsfonts}

\newcommand{\bbC}{\mathbb C}

\newcommand{\bfa}{{\mathbold a}}
\newcommand{\bfb}{{\mathbold b}}
\newcommand{\bfc}{{\mathbold c}}
\newcommand{\bfd}{{\mathbold d}}
\newcommand{\bfe}{{\mathbold e}}
\newcommand{\bff}{{\mathbold f}}

\newcommand{\bfm}{{\mathbold m}}
\newcommand{\bfn}{{\mathbold n}}

\newcommand{\bfv}{{\mathbold v}}

\newcommand{\bfx}{{\mathbold x}}

\newcommand{\bfA}{{\mathbold A}}
\newcommand{\bfB}{{\mathbold B}}
\newcommand{\bfC}{{\mathbold C}}

\newcommand{\bfE}{{\mathbold E}}
\newcommand{\bfF}{{\mathbold F}}

\newcommand{\bfI}{{\mathbold I}}

\newcommand{\bfL}{{\mathbold L}}

\newcommand{\bfT}{{\mathbold T}}

\newcommand{\bfV}{{\mathbold V}}
\newcommand{\bfW}{{\mathbold W}}
\newcommand{\bfX}{{\mathbold X}}

\newcommand{\beq}{\begin{equation}}
\newcommand{\eeq}{\end{equation}}
\newcommand{\beqs}{\begin{eqnarray}}
\newcommand{\eeqs}{\end{eqnarray}}
\newcommand{\beql}{\begin{equation} \label}

\newcommand{\bfchi}{\mathbold{\chi}}

\newcommand{\bfalpha}{\mathbold{\alpha}}

\newcommand{\bfbeta}{\mathbold{\beta}}

\newcommand{\bfPhi}{\mathbold{\Phi}}

\newcommand{\mTheta}{\mathnormal{\Theta}}
\newcommand{\mOmega}{\mathnormal{\varOmega}}
\newcommand{\mGamma}{\mathnormal{\Gamma}}
\newcommand{\mLambda}{\mathnormal{\Lambda}}

\newcommand{\bfzero}{\mathbf{0}}

\newcommand{\parderiv}[2]{\frac{\partial #1}{\partial #2}}

\newcommand{\mdeg}[1]{{#1}^{\circ}}

\usepackage[linktoc=all]{hyperref}
\hypersetup{colorlinks=true,
            linkcolor=blue,
            citecolor=red
}

\usepackage[margin=1in]{geometry}
\newcommand{\mUpsilon}{\mathnormal{\Upsilon}}

\usepackage{titlesec}

\titleformat{\paragraph}
{\normalfont\normalsize\bfseries}{\theparagraph}{1em}{}
\titlespacing*{\paragraph}
{0pt}{3.25ex plus 1ex minus .2ex}{1.5ex plus .2ex}

\PassOptionsToPackage{normalem}{ulem}
\usepackage{ulem}

\providecolor{added}{rgb}{0,0,1}
\providecolor{deleted}{rgb}{1,0,0}

\newcommand{\mum}[1]{(#1\,\mu m)^2}

\usepackage[symbol]{footmisc}

\begin{document}

\pagenumbering{roman}

\title{Dislocation pattern formation in finite deformation crystal plasticity}
\author{Rajat Arora\thanks{Dept.~of Civil \& Environmental Engineering, Carnegie Mellon University, Pittsburgh, PA 15213, email: rarora1@andrew.cmu.edu.}   \qquad Amit Acharya\thanks{Dept.~of Civil \& Environmental Engineering, and Center for Nonlinear Analysis, Carnegie Mellon University,
Pittsburgh, PA 15213, email: acharyaamit@cmu.edu.}
}
\date{November 3, 2018}

\maketitle

\begin{abstract} 
\noindent Stressed dislocation pattern formation in crystal plasticity at finite deformation is demonstrated for the first time. Size effects are also demonstrated within the same mathematical model. The model involves two extra material parameters beyond the requirements of standard classical crystal plasticity theory. The dislocation microstructures shown are decoupled from  deformation microstructures, and emerge without any consideration of latent hardening or constitutive assumptions related to cross-slip. Crystal orientation effects on the pattern formation and mechanical response are also demonstrated. The manifest irrelevance of the necessity of a multiplicative decomposition of the deformation gradient, a plastic distortion tensor, and the choice of a reference configuration in our model to describe the micromechanics of plasticity as it arises from the existence and motion of dislocations is worthy of note.

\end{abstract}

\pagenumbering{arabic}

\section{Introduction}\label{sec:overview}
Plastic deformation in crystals arises mainly due to the motion of dislocations under the action of externally applied stresses. The mutual interaction of dislocations under applied loads leads  to the development of intricate dislocation patterns such as dislocation cells \cite{mughrabi1976asymmetry,mughrabi1979persistent,mughrabi1981cyclic, hughes2000microstructure}  and labyrinths \cite{jin1984dislocation}, often with dipolar dislocation walls, and mosaics \cite{theyssier1995mosaic}. These microstructures appear at mesoscopic length scales in between the atomic and macroscopic scales. It is a fundamental challenge of theories and models of plasticity to predict such microstructure, with the attendant, often large, deformation and internal stress fields.

Different approaches have been used in the literature to model the development of dislocation microstructures such as \cite{ortiz1999nonconvex, limkumnerd2006mesoscale, chen2010bending, xia2015computational}, and other references mentioned therein. In the work of Ortiz and Repetto \cite{ortiz1999nonconvex}, dislocation structures at finite deformation have been shown to be compatible with  deformation fields that are minimizers of a pseudoelastic energy functional for a discrete time step of a rate independent crystal plasticity formulation. The predicted dislocation microstructures are necessarily stress-free by construction with non-dipolar walls (i.e., walls with non-zero net Burgers vector), and are accompanied by slip-band deformation microstructures. A key ingredient  for obtaining both the deformation and dislocation microstructures is the non-convex nature of the incremental energy functional, which in turn is the outcome of the use of strong latent hardening promoting local single-slip in their model.  

Sethna and co-workers \cite{limkumnerd2006mesoscale, chen2010bending} demonstrate (non-dipolar) dislocation walls with and without the presence of dislocation climb, showing the formation of self-similar dislocation microstructure starting from smooth random initial conditions. Their model is `minimal' in nature, involving geometrically linear kinematics for the displacement field, and a transport equation for the Nye tensor density \cite{nye1953} field arising from a conservation statement for the Burgers vector. On the other hand, Xia and El-Azab \cite{xia2015computational} demonstrate dislocation microstructure as an outcome of a model that assumes geometrically linear kinematics for the total deformation coupled to a system of stress-dependent, nonlinear transport equations for vector-valued slip-system dislocation densities. These slip system density transport equations involve complicated constitutive assumptions related to cross-slip, and the authors promote the point of view that dislocation patterning is necessarily related to the modeling of dislocation density transport at the level of slip system densities and the modeling of cross-slip.

The emergence of spatial inhomogeneity in the Nye tensor field was also reported in \cite{roy2006size, puri2011mechanical} at small deformations, utilizing a model referred to as Mesoscale Field Dislocation Mechanics (MFDM), that encompasses those used in \cite{limkumnerd2006mesoscale, chen2010bending}. In particular, these latter works do not account for `statistical dislocations', those that are responsible for most of the plastic deformation at the length scales in question where individual dislocations are not resolved (MFDM accounts for such). The model in \cite{xia2015computational} belongs to the same mathematical class as MFDM, being physically more involved with more state descriptors and associated coupled, nonlinear, equations of evolution. An attempt to understand the emergence of microstructure in this collection of models was made in \cite{roy2006size,das2016microstructure}, in drastically simplified 1-d settings. The conclusion in \cite{das2016microstructure} was that in all likelihood such complexity is not essential for the emergence of dislocation microstructure in this family of models; the nature of the fundamental transport equation for Nye tensor evolution coupled to stress along with the simplest representations, from conventional plasticity theory, of the plastic strain rate due to statistical dislocations, is adequate for the stated purpose, while being faithful to representing the plastic strain rate of both resolved and unresolved dislocation populations.

In this paper, we demonstrate that the aforementioned expectation is borne out in a full-fledged, geometrically nonlinear model of crystal plasticity based on MFDM. We demonstrate intricate spatial patterning, crystal orientation and size effects \cite{fleck1994strain,liu2012size, stelmashenko1993microindentations, ebeling1966dispersion}, the occurrence of stressed dislocation microstructures both under applied loads and in unloaded bodies, all in a rate-dependent setting with the simplest possible isotropic model of work hardening, relying in no way on non-convexity of any energy functional, incremental or otherwise.

In closing this brief review of related approaches we mention the Continuum Dislocation Dynamics framework of Hochrainer and collaborators; \cite{hochrainer2007three, hochrainer2016thermodynamically, sandfeld2015pattern} are some representative works. These models are developed based on a kinetic theory like framework, starting from the assumption that a fundamental statement for the evolution of a number density function on the space of dislocation segment positions and orientations is available (which is in itself a non-closed statement even if one knows completely the rules of physical evolution of individual dislocations segments of connected lines). Also, what a \emph{number} density of dislocations is supposed to mean for a tangled web of dislocation curves in a 3-d volume is not clarified. On making various assumptions for tractability, the theory produces (non-closed) statements of evolution for the averaged dislocation density (akin to the mesoscale Nye tensor field), the total dislocation density (similar to an appropriate sum of the averaged Nye tensor density and the Statistical density) and, these densities being defined as physical scalars, an associated curvature density field. Closure assumptions are made to cut off infinite hierarchies, which is standard for averaging based on nonlinear `microscopic equations', and further closure assumptions for constitutive statements are made based on standard thermodynamic arguments \cite{hochrainer2016thermodynamically}. The basic framework does not account for exact geometrically nonlinear continuum mechanics of deformation and stresses appropriate for large deformation plasticity. The models have been primarily exercised in situations involving a single slip plane. The work in \cite{sandfeld2015pattern} demonstrates some `patterning' in a simplified 2-d setting where total density concentrates (by approximately $4$ times) in `blobs' (terminology of the authors) covering most of the domain, with low densities restricted to narrow `walls', which is an inversion of what is observed in dislocation cells where high dipolar densities concentrate in narrow walls, with low densities (by orders of magnitude) arising in cell interiors.

We also note the finite deformation discrete dislocation plasticity formulation presented in the works of  \cite{deshpande2003finite, irani2015finite}. The latter work attempts to address the violation of the hypoelastic constitutive equation for stress of the dislocation fields in the computational implementation of the model proposed in \cite{deshpande2003finite}. Both formulations rely heavily on the superposition of linear elastic stress fields of individual dislocations (which seems counter-intuitive in the nonlinear setting, even for small elastic strain). Unfortunately, we have found the formulation in \cite{irani2015finite} to be not entirely transparent, thus hindering our understanding of the basic theory that is computationally implemented (compounded with typographical errors, e.g.~equations (16) and (17) therein that are important to understanding the computation of their $F^e$ tensor). For example, to what extent a constitutive statement like equation (32a) therein is an appropriate representation of frame-indifferent hyperelastic response, and better than the criticism leveled by the authors against the use of the (Jaumann rate-based) hypoelastic stress response proposed in \cite{deshpande2003finite}, is not clear to us. Clearly, the form of the strain measure utilized in equation (32a) suggests the use of linearised elasticity out of the current configuration, and then why the classical elastic solutions for dislocations from linear elasticity should be correct for linearised elasticity out of a configuration with \emph{stress} is not clarified - as is well-understood, the equations for linear elasticity and linearised elasticity differ when the configuration on which the problems are solved is under stress, leading to important nonlinear geometric effects like buckling instabilities.

This paper is organized as follows: Section \ref{sec:notation} reviews the notation and terminology used in the paper. Section \ref{sec:theoritical_framework} gives a brief introduction to the governing equations of finite deformation Mesoscale Field Dislocation Mechanics. The numerical algorithm used for computing approximate solutions and brief details of the finite element discretization of the equations of finite deformation MFDM  are presented in Section \ref{sec:fem}. Section \ref{sec:se_res_crystal} demonstrates the results obtained by using the developed computational framework.  Finally, some concluding remarks are presented in Section \ref{sec:conclusion}.

\section{Notation and terminology}
\label{sec:notation}
Vectors and tensors  are represented by bold face lower and upper-case letters, respectively. The action of a second order tensor $\bfA$ on a vector $\bfb$ is denoted by $\bfA\bfb$. The inner product of two vectors is denoted by $\bfa\cdot\bfb$ and the inner product of two second order tensors is denoted by $\bfA:\bfB$. A superposed dot denotes a material time derivative. A rectangular Cartesian coordinate system is invoked for ambient space and all (vector) tensor components are expressed with respect to the basis of this coordinate system. $(\cdot)_{,i}$ denotes the partial derivative of the quantity $(\cdot)$ w.r.t.~the $x_i$ coordinate direction of this coordinate system. Einstein's summation convention is always implied unless mentioned otherwise. The condition that any quantity (scalar, vector, or tensor) $a$ is defined to be $b$ is indicated by the statement $a := b$ (or $b =: a$). The symbol $|(\cdot)|$ represents the magnitude of the quantity $(\cdot)$.

The symbols $grad$, $div$, and $curl$ represent the gradient, divergence, and curl on the current configuration.  For a second order tensor $\bfA$, vectors $\bfv$, $\bfa$, and $\bfc$, and a spatially constant vector field $\bfb$, the operations of $div$, $curl$, and cross product of a tensor with a vector ($\times$) are defined as follows:
\begin{align*}
(div\bfA)\cdot\bfb &= div(\bfA^T \bfb), ~~~~~~~~~ \forall ~ \bfb \\
\bfb\cdot(curl\bfA)\bfc &=  \left[curl(\bfA^T \bfb)\right] \bfc, ~~~ \forall ~ \bfb, \bfc \\
\bfc\cdot(\bfA\times\bfv)\bfa &= \left[(\bfA^T \bfc)\times \bfv \right]\bfa ~~~~\forall ~ \bfa, \bfc.
\end{align*}
In rectangular Cartesian coordinates, these are denoted by
\begin{align*}
(div\bfA)_i =  A_{ij,j},\\
(curl\bfA)_{ri} =  \varepsilon_{ipq}A_{rq,p}, \\
(\bfA \times \bfv)_{ri} = \varepsilon_{ipq}A_{rp}v_q,
\end{align*} where $\varepsilon_{ijk}$ are the components of the third order alternating tensor $\bfX$. $\bfI$ is the second order Identity tensor whose components w.r.t.~any orthonormal basis are denoted by $\delta_{ij}$. The vector $\bfX(\bfA\bfB)$ is defined by
\begin{align*}
\left[\bfX(\bfA\bfB)\right]_i =  \varepsilon_{ijk}A_{jr}B_{rk}.
\end{align*}

In this paper, we qualitatively define \emph{patterning}  as the appearance of inhomogeneous distributions of dislocation density, more or less in the entire domain.

\section{Theory}
\label{sec:theoritical_framework}
This section presents a brief description of the governing equations and the initial and boundary conditions of finite deformation (Mesoscale) Field Dislocation Mechanics theory. Field Dislocation Mechanics (FDM) was developed in \cite{acharya2001model, acharya2003driving, acharya2004constitutive} building on the pioneering works of Kr\"oner \cite{kroner1981continuum}, Willis \cite{willis1967second},  Mura \cite{mura1963continuous}, and Fox \cite{fox1966continuum}. The theory utilizes a tensorial description of dislocation density \cite{nye1953, bilby1955continuous}, which is related to special gradients of the (inverse) elastic distortion field.  The governing equations of FDM at finite deformation are presented below:
\begin{subequations}
\begin{align}
&~\mathring{\bfalpha}\equiv (div\,\bfv)\bfalpha+\dot{\bfalpha}-\bfalpha\bfL^T = -curl\left(\bfalpha\times \bfV \right)
\label{eq:fdm_alpha}\\[1mm]
&~\bfW = \bfchi+grad\bff ; \quad\bfF^{e} := \bfW^{-1} \nonumber\\[1.25mm]
&\left.\begin{aligned}
& curl\bfW = curl{\bfchi} = -\bfalpha\\
&div{\bfchi} = \bf0
\label{eq:fdm_chi}
\end{aligned}~~~~\qquad\qquad\right\}\\[1.25mm]
&~div\left(grad\dot{\bff}\right) = div\left(\bfalpha\times \bfV - \dot{\bfchi}-\bfchi\bfL\right)\label{eq:fdm_fevol}\\
&~\rho\dot{\bfv} = div\,\bfT 
\label{eq:fdm_eqb}
\end{align}
\label{eq:fdm}
\end{subequations}
Here, $\bfF^e$ is the elastic distortion tensor, ${\bfchi}$ is the incompatible part of $\bfW$, $\bff$ is the plastic position vector \cite{roy2006size}, $grad\bff$ represents the compatible part of $\bfW$, $\bfalpha$ is the dislocation density tensor, $\bfv$ represents the material velocity field, $\bfL=grad\bfv$ is the velocity gradient,  $\bfT$ is the (symmetric) Cauchy stress tensor, and $\rho$ is the mass density. The dislocation velocity, $\bfV$, at any point is the instantaneous velocity of the dislocation complex at that point relative to the material; at the microscopic scale, the dislocation complex at most points consists of single segment with well-defined line direction and Burgers vector. At the same scale, the mathematical model assigns a single velocity to a dislocation junction, allowing for a systematic definition of a thermodynamic driving force on a dislocation complex that consistently reduces to well-accepted notions when the complex is a single segment, and which does not preclude dissociation of a junction on evolution.

The statement of dislocation density evolution \eqref{eq:fdm_alpha} is derived from the fact that the rate of change of Burgers vector content of any arbitrary area patch has to be equal to the flux of dislocation lines into the area patch carrying with them their corresponding Burgers vectors. Equation \eqref{eq:fdm_chi} is the fundamental statement of elastic incompatibility and relates the dislocation density field to the incompatible part of the inverse elastic distortion field $\bfW$. It can be derived by considering the closure failure of the image of every closed loop in the current configuration on mapping by $\bfW$. Equation \eqref{eq:fdm_fevol} gives the evolution equation for the compatible part of the inverse elastic distortion field. It can be shown to be related to the permanent deformation that arises due to dislocation motion \cite{acharya2004constitutive}. The field $grad \bff$ can also be viewed as the gradient of the inverse deformation for purely elastic deformations. Equation \eqref{eq:fdm_eqb} is the balance of linear momentum (in the absence of body forces). Balance of mass is assumed to hold in standard form, and balance of angular momentum is satisfied by adopting a symmetric stress tensor.

Equation \eqref{eq:fdm} is augmented with constitutive equations for the dislocation velocity $\bfV$ and the stress $\bfT$ in terms of $\bfW$ and $\bfalpha$ \cite{acharya2004constitutive, zhang2015single} to obtain a closed system. It can also be succinctly reformulated as
\begin{equation}\label{eqn:HJ}
\begin{split}
\dot{\bfW} & = -\bfW \bfL - (curl \bfW) \times \bfV\\
\rho \dot{\bfv} & = div \, \bfT
\end{split}
\end{equation} 
but since the \emph{system} of Hamilton-Jacobi equations in \eqref{eqn:HJ}$_1$ is somewhat daunting, we work with \eqref{eq:fdm} instead, using a Stokes-Helmholtz decomposition of the field $\bfW$ and the evolution equation for $\bfalpha$ in the form of a conservation law.

FDM is a model for the representation of dislocation mechanics at a scale where individual dislocations are resolved. In order to develop a model of plasticity that is applicable to mesoscopic scales, a space-time averaging filter is applied to microscopic FDM  \cite{acharya2006size, acharya2011microcanonical, babic1997average} and the resulting averaged model is called Mesoscale Field Dislocation Mechanics (MFDM). 
For any microscopic field $m$, the weighted, space-time running average field $\overline{m}$ is given as
\begin{align*}
\overline{m}(\bfx, t) := \dfrac{1}{\int_{B(\bfx)} \int_{I(t)} w(\bfx-\bfx', t-t')d\bfx' dt'  } {\int_{\mLambda} \int_{\mOmega} w(\bfx-\bfx', t-t') \,m(\bfx',t') d\bfx' dt'},
\end{align*} where $\mOmega$ is the body and $\mLambda$ is a sufficiently large interval of time. $B(\bfx)$ is a bounded region within the body around the point $\bfx$ with linear dimension of the spatial resolution of the model to be developed, and $I(t)$ is a  bounded interval contained in  $\mLambda$. The weighting function $w$ is non-dimensional and assumed to be  smooth in the variables $\bfx, \bfx', t, t'$. For fixed $\bfx$ and $t$, $w$ is only non-zero in $B(\bfx) \times I(t)$ when viewed as a function of $\bfx'$ and $t'$.

Assuming that all averages of products are equal to the product of averages except for $\overline{\bfalpha \times \bfV}$, the full set of governing equations of finite deformation MFDM theory (without inertia) can be written as 
\begin{subequations}
\begin{align}
&~\mathring{\overline\bfalpha}\equiv (div\,\overline\bfv)\overline\bfalpha+\dot{\overline\bfalpha}-\overline\bfalpha\overline\bfL^T = -curl\left(\overline\bfalpha\times \overline\bfV + \bfL^p\right)\label{eq:mfdm_alpha}\\[1mm]
&~\overline\bfW = \overline\bfchi+grad\overline\bff \nonumber\\[1.25mm]
&\left.\begin{aligned}
&curl{\overline{\bfW}} = curl{\overline\bfchi} = -\overline\bfalpha\\
&div{\overline\bfchi} = \bf0
\label{eq:mfdm_chi} 
\end{aligned}\right\}\\[1.25mm]
&~div\left(grad\dot{\overline\bff}\right) = div\left(\overline\bfalpha\times \overline\bfV + \bfL^p - \dot{\overline\bfchi}-\overline\bfchi\overline\bfL\right)\label{eq:mfdm_fevol}\\
&~div\,\overline\bfT  = \bfzero,
\label{eq:mfdm_eqb}
\end{align}
\label{eq:mfdm}
\end{subequations}
where $\bfL^p$ is defined as
\begin{align}\label{eqn:Lp}
\bfL^p(\bfx,t) := \overline{(\bfalpha - \overline{\bfalpha}(\bfx,t)) \times \bfV}(\bfx,t) = \overline{\bfalpha \times \bfV}(\bfx,t) - \overline{\bfalpha}(\bfx,t) \times \overline{\bfV}(\bfx,t).
\end{align}
The barred quantities in \eqref{eq:mfdm} are simply the weighted, space-time, running averages of their corresponding microscopic fields used in \eqref{eq:fdm}. The field $\overline{\bfalpha}$ is  the Excess Dislocation Density (ED). The microscopic density of Statistical Dislocations (SD)  at any point is defined as the difference between the microscopic dislocation density $\bfalpha$ and its  averaged field $\overline{\bfalpha}$:
\begin{equation*}
\bfbeta(\bfx,\bfx',t,t') = \bfalpha (\bfx',t') - \overline{\bfalpha}(\bfx,t),
\end{equation*}
which implies
\begin{align}
\label{eqn:tot_gnd_ssd}
\begin{split}
\rho_t &= \sqrt{\rho_g^2 + \rho_s^2}\\
\rho_t(\bfx, t) := \sqrt{ \overline{ \left( \dfrac{|\bfalpha  |}{b}\right)^2} (\bfx,t)} \ \ ; \  \rho_g(\bfx, t) &:=    \dfrac{|\overline{\bfalpha}(\bfx,t)|}{b} \ \  ; \ \ \rho_s(\bfx, t) := \sqrt{ \overline{ \left( \frac{ |\bfbeta|}{b} \right)^2 }(\bfx, t) },
\end{split}
\end{align}
with $b$ the magnitude of the Burgers vector of a dislocation in the material, $\rho_t$ the \emph{total dislocation density}, $\rho_g$ the magnitude of ED (commonly referred to as the geometrically necessary dislocation density), and $\rho_s$ is, up to a scaling constant, the root-mean-squared SD. We refer to $\rho_s$ as the scalar statistical dislocation density (\emph{ssd}). It is important to note that spatially unresolved dislocation loops below the scale of resolution of the averaged model do not contribute to the ED ($\overline\bfalpha$) on space time averaging of the microscopic dislocation density, due to sign cancellation. Thus, the magnitude of the ED is an inadequate approximation of the total dislocation density. Similarly, a consideration of `symmetric' expansion of unresolved dislocation loops shows that the plastic strain rate produced by SD, $\bfL^p$ \eqref{eqn:Lp}, is not accounted for in $\overline{\bfalpha} \times \overline{\bfV}$, and thus the latter is not a good approximation of the total averaged plastic strain rate $\overline{\bfalpha \times \bfV}$.

In MFDM, closure assumptions are made for the field $\bfL^p$ and the evolution of $\rho_s$, as is standard in most, if not all, averaged versions of nonlinear microscopic models, whether of real-space or kinetic theory type. As such, these closure assumptions can be improved as necessary (and increasingly larger systems of such a hierarchy of nonlinear pde can be formally written down for MFDM). In this paper, we adopt simple and familiar closure statements from (almost) classical crystal plasticity theory and probe the capabilities of the model that results. Following the works of Kocks, Mecking, and co-workers \cite{mecking1981kinetics, estrin1984unified} we describe the evolution of $\rho_s$ through a statement, instead, of evolution of material strength $g$ described by \eqref{eq:softening}; $\bfL^p$ is defined by \eqref{eq:Lp} following standard assumptions of crystal plasticity theory and thermodynamics. A significant part of the tensorial structure of \eqref{eq:Lp} can be justified by  elementary averaging considerations of dislocation motion on a family of slip planes under the action of their Peach-Koehler driving force \cite{acharya2012elementary}.

\emph{Henceforth, we drop the overhead bars for convenience in referring to averaged quantities, and we will only refer to the `macroscopic' fields given in \eqref{eq:mfdm}}. Also, $\bfalpha$ will be simply referred to as the dislocation density tensor. 
Since the system in \eqref{eq:mfdm} is not closed,  $\bfT$, $\bfL^p$, and $\bfV$ are to be constitutively specified response functions specific to materials.

As shown in \cite{acharya2015dislocation}, \eqref{eq:mfdm_alpha} and \eqref{eq:mfdm_chi} imply
\begin{equation}
\dot{\bfW} + \bfW\bfL = \bfalpha \times \bfV + \bfL^p
\label{eq:vel_grad}
\end{equation} up to the gradient of a vector field, which is re-written as
\begin{equation*}
\bfL = \dot{\bfF^{e}} {\bfF^{e-1}} + \bfF^{e}(\bfalpha\times\bfV+\bfL^p).
\end{equation*}
This can be interpreted as the decomposition of the velocity gradient into an elastic part, given by $\dot{\bfF^{e}} {\bfF^{e-1}}$, and a plastic part given by  $\bfF^{e}(\bfalpha\times\bfV+\bfL^p)$. The plastic part is defined by the motion of dislocations, both resolved and unresolved, on the current configuration and \emph{no notion of any pre-assigned reference configuration is needed}. Of  significance is also the fact that \emph{MFDM involves no notion of a plastic distortion tensor and yet produces (large) permanent deformation}.

\subsection{Constitutive equations for \texorpdfstring{$\bfT$}{T}, \texorpdfstring{$\bfL^p$}{Lp}, and \texorpdfstring{$\bfV$}{V}} 
\label{sec:dissipation}
MFDM requires constitutive statements for the stress $\bfT$, the dislocation velocity $\bfV$, and the plastic distortion rate $\bfL^p$. We make the model consistent with the minimal, but essential,  requirement of non-negative dissipation through these choices. For this we consider the mechanical dissipation $D$ which, in the presence of inertia and body forces, is defined as the difference between the power of the applied forces and the rate of change of the sum of the kinetic and free energies of the system:
\begin{align*}
D &= \int_{\partial\varOmega} \bfT\bfn\cdot\bfv \,dA + \int_{\varOmega} \bfb\cdot\bfv \,dV  - \dot{\overline{\int_{\varOmega}\rho \, (\psi + \dfrac{1}{2}\bfv\cdot\bfv)\, dV}},
\end{align*}
where $\psi$ is the specific Helmholtz free-energy of the system, and $\bfb$ is the body force. The Helmholtz free energy of the system per unit mass, $\psi$, is assumed to be the sum of the elastic energy $\phi(\bfW)$ density and a term $\mUpsilon(\bfalpha)$ that is a heuristic representation of the averaging of a microscopic core energy, up to the mesoscale:
\begin{align*}
\psi = \phi(\bfW) + \mUpsilon(\bfalpha).
\end{align*}
The elastic energy per unit mass is specified as 
\begin{align}
\begin{split}
&\phi(\bfW) = \dfrac{1}{2\rho^*} \bfE^e:\bbC:\bfE^e\\[1.5mm]
&\bfE^e = \dfrac{1}{2} (\bfC^e - \bfI); \quad  \bfC^e = \bfW^{-T}\bfW^{-1},
\end{split}
\label{eq:phi_W}
\end{align}
where $\rho^*$ is the mass density of the pure, unstretched elastic lattice,  and $\bbC$ is the fourth order elasticity tensor, assumed to be positive definite on the space of second order symmetric tensors.
$\mUpsilon(\bfalpha)$ is specified as
\begin{align*}
\mUpsilon(\bfalpha) = \dfrac{1}{2\rho^*}{\epsilon} \,\bfalpha:\bfalpha,
\end{align*}
where $\epsilon$ is a material constant that has dimensions of $stress \times length^2$. Using the balances of mass and linear momentum, the definition of $\mUpsilon(\bfalpha)$, and the evolution equations for $\bfW$ \eqref{eq:vel_grad} and $\bfalpha$ \eqref{eq:mfdm_alpha}, the dissipation can be expressed as
\begin{align}
\begin{split}
D &= \int_{\varOmega} \bfT : \bfL \,dV - \int_{\varOmega}\rho \, \dot{\overline{(\phi(\bfW) + \mUpsilon(\bfalpha))}}\, dV \\
&= \int_{\varOmega} \left[ \bfT  + \rho  \bfW^T \frac{\partial \phi}{\partial \bfW} \right]: \bfL \, dV  - \int_{\varOmega} \rho  \bfX\left[\left(\frac{\partial \phi}{\partial \bfW} \right)^T \bfalpha \right] \cdot \bfV  dV  - \int_{\varOmega} \rho \frac{\partial \phi}{\partial \bfW}  : \bfL^p \,  dV\\
& \qquad+ \dfrac{\epsilon}{\rho^*} \left[ \int_{\varOmega} \rho  \left(  (\bfalpha : \bfalpha) \bfI -    \bfalpha^T \bfalpha \right) : \bfL\, dV  + \int_{\varOmega} \rho  \bfX \left( \left[ curl \bfalpha \right]^T \bfalpha \right) \cdot \bfV  dV \right. \\
& \qquad \qquad  \qquad + \left. \int_{\varOmega}\underbrace{\rho \, curl \bfalpha  : \bfL^p} \,  dV  - \int_{\partial \varOmega}  \rho  \,  \bfalpha:\left( (\bfalpha \times \bfV + \bfL^p) \times \bfn \right) \, dA \right].
\end{split}
\label{eq:dissipation}
\end{align}
From the study of solutions to FDM it is known \cite{acharya2011equation, zhang2015single} that the core energy provides a crucial physical regularization and therefore we want to keep the simplest possible effect of it in MFDM, in the absence of rigorous information on the averaged structure of FDM. Based on the above terms in the dissipation, if we assume $\bfL^p$ to be in the direction of its driving force to ensure non-negative dissipation, then it can be observed that the presence of $curl\bfalpha$ in the driving force for $\bfL^p$ gives rise to a term, in the evolution equation \eqref{eq:mfdm_alpha} for  $\bfalpha$, of the form $-curl (curl \bfalpha)$ with a (possibly spatially varying) non-negative coefficient. This additional term behaves as a diffusive regularization by a standard identity of vector calculus and the fact that $div\bfalpha = \bf0$. Motivated by these considerations related to the dissipation, we make the following constitutive assumptions for $\bfT$, $\bfV$, and $\bfL^p$ in MFDM.

Ensuring no dissipation in purely elastic processes, the stress is given by
\begin{align*}
\bfT = -\rho\bfW^T\parderiv{\phi}{\bfW} ~&\Rightarrow ~\bfT = \bfF^e\left[\bbC:\bfE^e\right]\bfF^{eT}.
\end{align*}
The above expression for the Cauchy stress tensor tacitly assumes that $\frac{\rho}{\rho^*}$ is absorbed in the elastic moduli  $\bbC$, which is assumed to be spatially constant in this work.

Classical crystal plasticity assumes $\bfL^p$ to be a sum of slipping on prescribed slip systems (cf.~\cite{asaro1983micromechanics}). To augment this assumption with an additive term in $\epsilon\, curl\bfalpha$ as motivated above requires the introduction of a mobility coefficient with physical dimensions of $(stress \times time^{-1})$. In the absence of more detailed knowledge, simplicity demands that all dissipative processes be linked to a common time scale and we do not proliferate material parameters. Thus, we assume the stress scale in the mobility to be linked to the initial yield strength $g_0$, and its time scale to be linked to the reciprocal of the average slip system slipping rates. These assumptions result in the coefficient of $curl \bfalpha$ in $\bfL^p$ to be $\frac{\epsilon}{g_0} \frac{1}{n_{sl}} \sum_k^{n_{sl}}\hat\gamma^k$ (up to a factor $\frac{\rho}{\rho^{*}}$) and defining $l^2 := \frac{\epsilon}{g_0}$ we assume $\bfL^p$ to be given by
\begin{align}
\bfL^p &= \underbrace{\bfW\, \left(\sum_k^{n_{sl}} \hat{\gamma}^k \, \bfm^k\otimes\bfn^k \right)_{sym}}_{\hat{\bfL}^p} + \quad  \left( \dfrac{l^2}{n_{sl}} \sum_k^{n_{sl}} |\hat\gamma^k| \right) \,curl \bfalpha
\label{eq:Lp}
\end{align}
where
\begin{align}
\hat{\gamma}^k &= sgn(\tau^k)\, \hat{\gamma_0}^k\left(\frac{|\tau^k|}{g} \right)^{\frac{1}{m}}. \label{eq:gammadotk}
\end{align}
In the above, $(\cdot)_{sym}$ represents the symmetric part of $(\cdot)$, $\hat\gamma_0$ is a reference strain rate, $\hat{\gamma}^k$ represents the magnitude of SD slipping rate on the slip system $k$, $n_{sl}$ is the total number of slip systems, $sgn(\tau^k)$ denotes the sign of the scalar $\tau^k$, and $g$ is the material strength. The vectors $\bfm^k$ and $\bfn^k$ represent the slip direction and the slip plane normal for the $k^{th}$ slip system in the current configuration. These are given as
\begin{align}
\bfm^k &= {\bfF^e\bfm_0^k}  \nonumber \\
\bfn^k &= {{\bfF^e}^{-T}\bfn_0^k}, \nonumber
\end{align}
where $\bfm_0^k$ and $\bfn_0^k$ are the corresponding unstretched  unit vectors. The resolved shear stress $\tau^k$ on the $k^{th}$ slip system is defined as
\begin{equation}
\tau^k = \bfm^k\cdot\bfT\bfn^k. \nonumber
\end{equation}
The use of the symmetrization in the definition of $\hat{\bfL}^p$ is not standard, but found to be necessary, following \cite[Sec.~5.5]{puri2011mechanical}.

We mention here that the length scale $l$ is not responsible for producing enhanced size effects and microstructure in MFDM. Rather, the `smaller is harder' size effect becomes more pronounced as $l$ decreases since its presence reduces the magnitude of the $\bfalpha$ field and consequently reduces hardening \eqref{eq:softening}. It plays a role in the details of the microstructural patterns which is explored in Sec.~\ref{sec:role_of_l}, while not being responsible for their generation, as shown in Sec. \ref{sec:k_0}.

The direction of the dislocation velocity, $\bfd$, is given by
\begin{eqnarray}
\bfd = \bfb - \left(\bfb\cdot \frac{\bfa}{|\bfa|} \right) \frac{\bfa}{|\bfa|}
\label{eq:V_dir}
\end{eqnarray} 
(for motivation see \cite{acharya2006size, acharya2012elementary}) with
\begin{align}
T'_{ij} = T_{ij} - \dfrac{T_{mm}}{3}\delta_{ij}; ~~
b_i := \varepsilon_{ijk}T'_{jr}{F^e}_{rp}\alpha_{pk}; ~~
a_i := \dfrac{1}{3}T_{mm} \varepsilon_{ijk}{F^e}_{jp}\alpha_{pk}.
\label{eq:V_ab}
\end{align}
The dislocation velocity is then assumed to be
\begin{equation}
 \bfV = \zeta \frac{\bfd}{|\bfd|}
 \label{eq:V}
\end{equation}
with 
\begin{equation}
 \zeta =  \dfrac{\mu^2\,\eta^2\, b}{g^2 \,n_{sl}}\sum_k^{\,n_{sl}} |\hat{\gamma}^k|, 
  \label{eq:V_zeta}
\end{equation}
where $b$ is as in \eqref{eqn:tot_gnd_ssd}, $\mu$ is the shear modulus, and $\eta = \frac{1}{3}$ is a material parameter.
The strength of the material is evolved according to (cf. \cite{acharya2000grain,beaudoin2000consideration, acharya2006size})
\begin{equation}
\dot{g} = \left[ \frac{\mu^2\eta^2b}{2(g-g_0)}k_0 \left|\bfalpha\right|+ \mTheta_0 \left(\frac{g_s-g}{g_s-g_0}\right)\right]\left(\left|\bfF^e\bfalpha\times\bfV\right|+ \sum_k^{n_{sl}} |\hat{\gamma}^k|\right),
\label{eq:softening}
\end{equation} 
where $\mTheta_0$ is the Stage $2$ hardening rate, $k_0$ is a material constant, and $g_s$ is the saturation material strength. 

The material parameters ($g_0, g_s, \mu, \hat\gamma_0, m, \mTheta_0$) mentioned above are part of the constitutive structure of well-accepted models of classical plasticity theory. Our model requires $2$ unknown fitting parameters: $l$, $k_0$, with the latter characterizing the plastic flow resistance due to ED. The material strength defines the $ssd$ distribution (see \eqref{eqn:tot_gnd_ssd}) as
\begin{align}
\rho_s := \left( \frac{g}{\eta \mu b} \right)^2.
\label{eq:ssd}
\end{align}

We note that for these choices of $\bfT$, $\bfV$, and $\bfL^p$
\begin{align*}
\lim_{\epsilon \to 0} D &= \int_\varOmega  \zeta\dfrac{\bfd}{|\bfd|} \cdot \bfX[\bfT\bfF^e\bfalpha] \, dV + \int_{\mOmega} \sum_k^{n_{sl}} \tau^k \hat{\gamma}^k  \, dV\\
& \geq 0
\end{align*}
(assuming the multiplier of $\epsilon$ within the square parenthesis in \eqref{eq:dissipation} is bounded in the limit).

\subsection{Boundary conditions}
\label{sec:boundary_conditions}
The incompatibility equation \eqref{eq:mfdm_chi} admits a boundary condition of the form 
\begin{align*}
\bfchi\,\bfn = \bf0 ~ \rm{on} ~\partial\varOmega,
\end{align*}
where $\bfn$ is the outward unit normal on the outer boundary $\partial \varOmega$ of the current configuration $\varOmega$. Such a boundary condition ensures vanishing $\bfchi$ in the absence of a dislocation field. The equilibrium equation \eqref{eq:mfdm_eqb}  admits standard  admissible traction and/or displacement boundary conditions. The dislocation evolution equation  \eqref{eq:mfdm_alpha} admits a  `convective' boundary condition of the form $(\bfalpha \times \bfV + \hat{\bfL}^p) \times \bfn=\bfPhi$ where $\bfPhi$ is a second order tensor valued function of time and position  on the boundary characterizing the flux of dislocations at the surface with unit normal field $\bfn$ satisfying the constraint $\bfPhi\bfn=\bf0$. A no slip or plastically constrained boundary condition is modeled by assuming $\bfPhi \equiv \bfzero$. We will also sometimes use a less restrictive boundary condition where we simply evaluate $\hat\bfL^p\times\bfn$ on the boundary (akin to an outflow condition) along with the specification of $\bfalpha (\bfV\cdot\bfn)$ on the inflow parts of the boundary (where $\bfV\cdot\bfn < 0$). This is referred to as the unconstrained case since dislocations are free to exit the domain without any added specification. In addition to this, for non zero $l$ Eq.~\eqref{eq:mfdm_alpha} also requires specification of $(\frac{l^2}{n_{sl}} \sum_k^{n_{sl}} |\hat\gamma^k|) curl\bfalpha \times \bfn$ on the boundary. For this work, we assume the input flux $\bfalpha(\bfV\cdot\bfn)$ and $curl\bfalpha\times\bfn$ to vanish on the boundary.  The evolution equation \eqref{eq:mfdm_fevol} for $\bff$ uses a Neumann boundary condition of the form 
\begin{align*}
(grad \dot{\bff})\bfn = \left(\bfalpha \times \bfV + \bfL^p - \dot{\bfchi} - \bfchi\bfL\right)\,\bfn.
\end{align*}

\subsection{Initial conditions}
The evolution equations for the dislocation density and $\bff$  (\eqref{eq:mfdm_alpha} and \eqref{eq:mfdm_fevol} respectively) require specification of initial conditions on the domain. The initial condition for \eqref{eq:mfdm_alpha} can be prescribed in the following form:
$\bfalpha(\bfx, t=0) = \bfalpha_0$. For this work, we take $\bfalpha_0 = \bf0$. The initial condition for \eqref{eq:mfdm_fevol} is given as the solution of 
\begin{align*}
curl\bfchi &= -\bfalpha_0\\
div\bfchi &= \bf0\\
div \bfT(\bfW) &= \bf0
\end{align*}
along with the specification of statically admissible traction boundary conditions. This corresponds to the determination of $\bff$, $\bfchi$, and stresses at $t = 0$ for a given dislocation density distribution on the initial configuration, i.e., the current configuration at $t = 0$, and will be referred to as the ECDD solve.  An auxiliary condition of $\dot{\bff} = \bf0$ at a point is needed to uniquely evolve $\bff$ from \eqref{eq:mfdm_fevol}. 

\section{Numerical implementation} 
\label{sec:fem}
The  finite element implementation of the system of equations given in \eqref{eq:mfdm} has been discussed in \cite{arora_xia_acharya_cmame} where detailed numerical algorithms, verification, and validation exercises are provided. Here, we briefly describe the general flow of the algorithm for the sake of being self-contained.

Along with the system of equations \eqref{eq:mfdm}$_{a-c}$, we solve the rate form of the equilibrium equation to obtain the material velocity field $\bfv$, which is used to obtain the discrete motion of the body. However, this may not satisfy (discrete) balance of forces at each time step. Therefore, we use the equilibrium equation \eqref{eq:mfdm_eqb} to correct for force balance in alternate time increments. In the absence of body forces and inertia, the equilibrium equation \eqref{eq:mfdm_eqb} in rate form is \cite{hill1959some,mcmeeking1975finite}
\begin{align}
div\left(div(\bfv)\,\bfT + \dot\bfT - \bfT\,\bfL^T\right) = \bf0.
\label{eq:mfdm_eqb_rate}
\end{align} This requires specification of velocity and/or statically admissible nominal traction rates on complementary parts of the boundary at all times.
Finite Element Method based computational modeling using MFDM requires the concurrent solution of the system of equations in \eqref{eq:mfdm} along with Eq.~\eqref{eq:mfdm_eqb_rate} resulting in $10$ degrees of freedom (DOFs) per node in $2$D. This includes $2$ unknowns in $\bfalpha$ ($\alpha_{13}$ and $\alpha_{23}$), $4$ in $\bfchi$ ($\chi_{11}, \chi_{12}, \chi_{21},$ and $\chi_{22}$), and $2$ each in $\bfv$ and $\bff$ respectively. The details of the staggered numerical implementation are discussed in  \cite{arora_xia_acharya_cmame}, which utilizes the following numerical schemes: Galerkin FEM for the equilibrium equation \eqref{eq:mfdm_eqb}, the rate form of the equilibrium equation \eqref{eq:mfdm_eqb_rate}, and compatible part of the inverse of elastic distortion \eqref{eq:mfdm_fevol};  Least-squares FEM \cite{jiang2013least} for the incompatibility equation \eqref{eq:mfdm_chi}; and Galerkin-Least-Squares FEM \cite{hughes1989new} for the dislocation evolution equation \eqref{eq:mfdm_alpha}. 

The numerical scheme presented in \cite{arora_xia_acharya_cmame} is independent of the constitutive assumptions made for $\bfL^p$ and $\bfV$; here, we use the specifications in Eqs.~\eqref{eq:Lp} and \eqref{eq:V}, respectively.

\subsection{Algorithm}
\label{sec:algorithm}
The system of equations \eqref{eq:mfdm} is solved by discretely evolving in time. A combination of explicit-implicit schemes have been chosen to evolve the system variables in time (cf. \cite{roy2006size}). An efficient time stepping criteria based on plastic relaxation, and purely elastic and `yield strain' related physical model parameters, has been defined. A `cutback' algorithm has been designed and is used to ensure a stable, robust, and accurate evolution of plastic response. The algorithm for solving the system of equations \eqref{eq:mfdm} is as follows:
\begin{enumerate}
\item Given the material parameters and initial condition on $\bfalpha$, ECDD is solved to specify $\bff(t=0)$, $\bfchi(t=0)$ and the initial stresses on the configuration at time $t=0$.

In any given time step $[t^n, t^{n+1}]$ with the current configuration and state known at time $t^n$ and with $(\cdot)^n$ respresenting the quantity $(\cdot)$ at time $t^{n}$,

\item The rate form \eqref{eq:mfdm_eqb_rate} of the equilibrium equation \eqref{eq:mfdm_eqb} is solved on the configuration at $t^n$ to obtain the material velocity field $\bfv$ in the interval $[t^n, t^{n+1}]$. The velocity field is used to obtain the current configuration at time $t^{n+1}$.

\item $\bfalpha$ is evolved from \eqref{eq:mfdm_alpha} on the configuration at time $t^n$ to define the dislocation density field, $\bfalpha^{n+1}$, on the configuration at time $t^{n+1}$.
\item  $\bfchi^{n+1}$  is defined on the configuration at time $t^{n+1}$ by solving \eqref{eq:mfdm_chi} on the same configuration with $\bfalpha^{n+1}$ as data.

\item \label{algo_force_calculation} The nodal (reaction) forces on the part of the boundary with specified boundary conditions on material velocity are evaluated as follows: assume the nodal forces are known at  time $t^n$. On solving Eq.~\eqref{eq:mfdm_eqb_rate} on the configuration at $t^n$, a (reaction) nodal force rate field on the velocity-Dirichlet boundary is generated. For each node on this part of the boundary, this reaction force rate physically corresponds to the spatial integration of the nominal/First Piola Kirchhoff traction rate, based on the configuration at time $t^n$ as reference, over the area patch (on the same configuration) that contributes to the node in question. Since such a nodal force rate, viewed as a discrete function of time, corresponds to the evolving current configuration of the body (recall the definition of the First Piola-Kirchhoff stress tensor), we simply (discretely) integrate it in time and accumulate the result on the known nodal force at time $t^n$ to obtain the nodal force (on the velocity-Dirichlet-part of the boundary) at time $t^{n+1}$.

\item $\bff^{n+1}$ is determined as follows:

\begin{itemize}

\item The evolution equation \eqref{eq:mfdm_fevol} for $\bff$ is solved on the configuration at time $t^n$ to define $\bff^{n+1}$ on the configuration at time $t^{n+1}$.

\item In alternate increments, the  equilibrium equation   \eqref{eq:mfdm_eqb}  is solved on the configuration at time $t^{n+1}$ for the field $\bff$, in order to satisfy balance of forces. The problem is posed as a traction boundary value problem with nodal forces calculated in step \ref{algo_force_calculation} above. $\bff^{n+1}$ obtained by solving the evolution equation \eqref{eq:mfdm_fevol}  serves as the guess for the   Newton-Raphson based scheme.

\end{itemize}

\end{enumerate}

  Conventional plasticity theories do not account for the plastic strain rate of the (excess) dislocation motion nor the boundary conditions related to ED flow at the boundaries of the body. In  MFDM, we can recover  classical plasticity theory by setting $\bfV = \bfzero$ and $l = 0$ in the system given in \eqref{eq:mfdm}, $k_0= 0$ in \eqref{eq:softening}, and treating the external boundary as plastically unconstrained as mentioned in Sec.~\ref{sec:boundary_conditions}.

MFDM may be viewed as a thermodynamically consistent strain gradient plasticity theory without higher order stresses.

\section{Results}
\label{sec:se_res_crystal}
The formulation presented in Sec.~\ref{sec:fem} above  is implemented in a C$++$  code based  on the deal.ii \cite{dealII85} framework to carry out finite-element computations. Bilinear elements for employed to approximate all fields.

Sec.~\ref{sec:multiple_slip} focuses on the case when multiple slip systems are present in the body. We demonstrate size effects and the emergence of dislocation patterns and dipolar dislocation walls. We also study the effect of orientation on the microstructural patterns and the macroscopic stress-strain response of the material. In Sec.~\ref{sec:single_slip} we look at the microstructural patterning and effect of orientation for the special case when only one slip system is present. Sec.~\ref{sec:convergence} briefly discusses the convergence of microstructural patterns and overall stress-strain response with respect to mesh refinement. Sec.~\ref{sec:k_0} presents a necessary condition for pattern formation in the model. Finally, Sec.~\ref{sec:role_of_l} focuses on the effect of the length scale $l$ on microstructural patterns.

\begin{figure}[H]
    \centering
{\includegraphics[width=.7\linewidth]{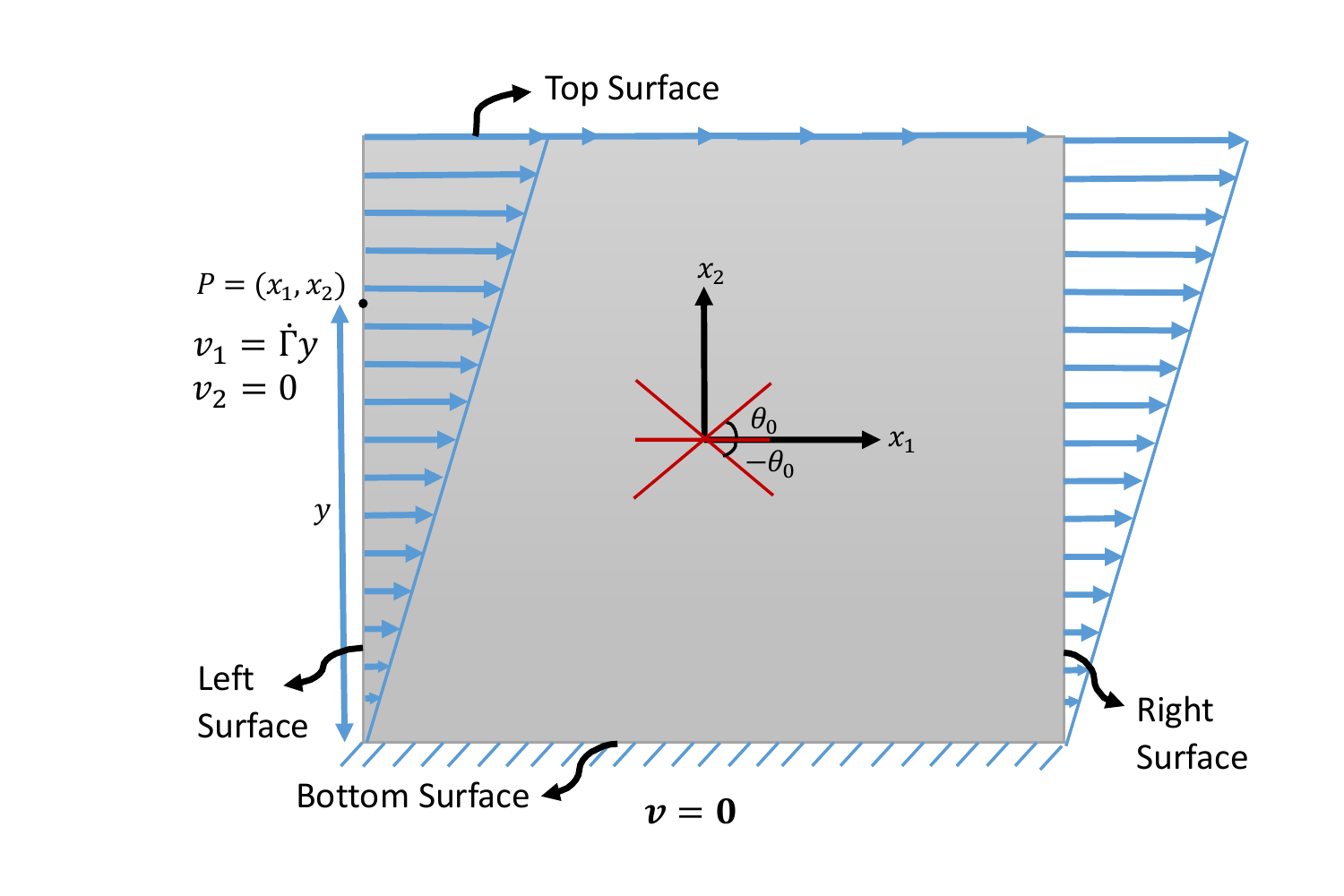}}
        \caption{Schematic layout of a typical model geometry.}
            \label{fig:se_schematic}
\end{figure}

Simulations are performed on square domains of sizes $\mum{1}$, $\mum{5}$, and $\mum{100}$. The details of the meshes employed are given in Table \ref{tab:mesh} with rationale presented in Sec.~\ref{sec:convergence}. Velocity b.c.s~corresponding to overall simple shear are imposed for a plane strain problem. At any point $P= (x_1, x_2)$ on the boundary a velocity of $v_2 = 0$ and $v_{1} = \hat{\mGamma} y(x_2)$ is imposed, where $y(x_2)$ is the height of the point $P$ from the bottom surface as shown in the schematic of the problem in Figure \ref{fig:se_schematic}. $\hat{\mGamma}$ is the applied shear strain rate. The initial ($t = 0$) slip system orientation will be denoted by the parameter $\theta_0$.

\begin{table}[htbp]
        \centering
        \begin{tabular}{ c  r  }
\hline
Sample Size &   Mesh \\ \hline
 $\mum{1}$ & $70 \times 70$ \\ 
 $\mum{5}$ & $70 \times 70$ \\ 
 $\mum{100}$ & $70 \times 70$\\ \hline
\end{tabular}
\caption{Details of finite element mesh used in computations.}
\label{tab:mesh}
\end{table}

The conventional plasticity solution plotted in the figures below is obtained by numerically integrating the evolution equation for the elastic distortion tensor $\bfF^e$ given by \eqref{eq:conventional_Fedot} to obtain the Cauchy stress response for an imposed spatially homogeneous velocity gradient history, $\bfL$, corresponding to a simple shearing motion:
\begin{align}
\begin{split}
& \dot\bfF^e = \bfL\bfF^e - \bfF^e\bfL^p\bfF^e =: \tilde\bff(\bfF^e, g)\\
&\dot g = \tilde{g}(\bfF^e, g)
\end{split}
\label{eq:conventional_Fedot}
\end{align}
where $\bfL^p$ is defined from \eqref{eq:Lp} with $l = 0$, and $\tilde{g}$ is given by \eqref{eq:softening} with $k_0 = 0$. 

\begin{table}[htbp]
        \centering
        \begin{tabular}{ c  r  }
\hline
Parameter &   Value  \\ \hline
 $b$ & $4.05 $\AA  \\ 
 $g_0$ & $17.3$ MPa  \\ 
 $g_s$ &  $161$ MPa \\ 
 $\mTheta_0$ &  $392.5$ MPa \\ 
 $m$ &  $.03$ \\         
 $E$ &  $62.78$ GPa \\     
 $\nu$ &  $.3647$ \\ 
 $\hat\mGamma$ &  $1\, s^{-1}$ \\ 
 $\hat\gamma_0$ & $1\,s^{-1}$  \\ \hline    
 $k_0$ & $20$  \\ 
 $l$ & $\sqrt{3} \times 0.1 \,\mu m$  \\\hline 
\end{tabular}
\caption{Default parameter values used in computations.}
\label{tab:se_parameters}
\end{table}



The stress-strain behavior of the body is modeled by plotting the averaged $T_{12}$ on the top surface which is denoted by $\tau$ in the subsequent figures. $\tau$ is calculated by  summing the tangential components  of the nodal reaction force on the top surface and then dividing by the current area (line length) of the surface. The strain $\mGamma$ at any time $t$ is given as $\mGamma = \hat{\mGamma}\,t$. All material parameters used in the simulations are presented in Table \ref{tab:se_parameters}. $E$ and $\nu$ denote the Young's modulus and Poisson's ratio, respectively.


\subsection{Dislocation microstructure and size effect in multiple slip}
\label{sec:multiple_slip}
For multiple slip, we assume that there are $3$ slip systems present in the crystal, oriented at $\mdeg{\theta_0}, \mdeg{-\theta_0}$, and $\mdeg{0}$ from the $x$ axis as shown above in Fig.~\ref{fig:se_schematic}. The slip directions and normals for the $3$ slip systems are given as
\begin{align*}
&\bfm_0^1 = (\cos(\theta_0), \sin(\theta_0))  &\bfn_0^1 = (-\sin(\theta_0), \cos(\theta_0))\\
&\bfm_0^2 = (\cos(0), \sin(0)) 
&\bfn_0^2 = (-\sin(0), \cos(0))\\
&\bfm_0^3 = (\cos(\theta_0), -\sin(\theta_0)) &\bfn_0^3 = (\sin(\theta_0), \cos(\theta_0)).
\end{align*} 
Thus, $\theta_0$ characterizes the orientation of all the slip systems at $t = 0$. 

\subsubsection{Size effect}
We demonstrate size effects in elastic-plastic material behavior up to large strains for both the plastically constrained and unconstrained cases defined in Section \ref{sec:boundary_conditions}.

\begin{figure}[H]
        \centering
{\includegraphics[width=.7\linewidth]{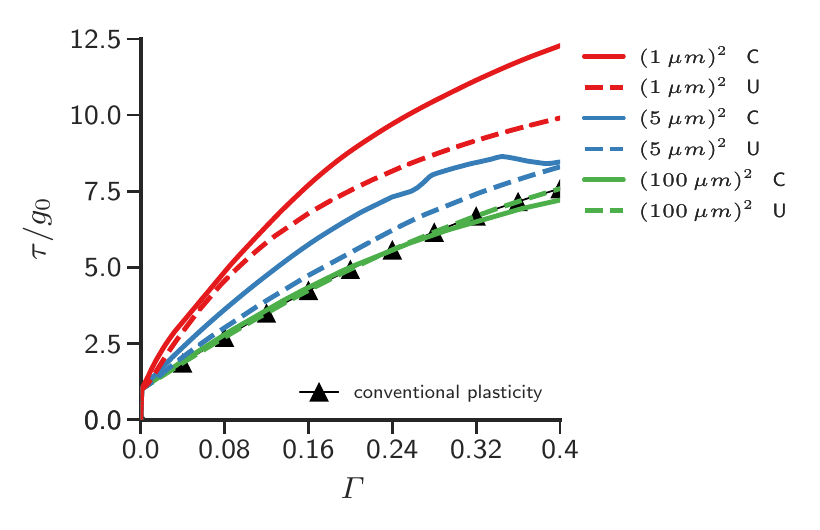}}
        \caption{Size effect under simple shear C: Constrained Boundaries U: Unconstrained Boundaries.}
    \label{fig:se_ss}
 \end{figure}

Figure \ref{fig:se_ss} shows the averaged stress-strain ($\tau$ vs.~$\mGamma$) response for all the domain sizes and both boundary conditions (plastically constrained and unconstrained), demonstrating the `smaller is harder' size effect under simple shear. These results are  in qualitative agreement with experimental observations \cite{fleck1994strain,liu2012size, stelmashenko1993microindentations, ebeling1966dispersion}. 

For the unconstrained case, the response of the larger domain size  of $\mum{100}$ overlaps the conventional plasticity solution. This is expected as the larger sample develops no inhomogeneities in deformation and therefore $|\bfalpha| \approx 0$.  However, the smaller domain sizes $\mum{1}$ and $\mum{5}$ develop inhomogeneity at small strains of (approximately) $0.5\%$ and $1.7\%$, respectively. This (controlled) instability leads to deviation from the homogenous solution, which in turn increases the local hardening, resulting in harder response than the conventional solution. This instability of the time-dependent spatially homogeneous simple shearing solution for the MFDM theory is discussed in \cite{roy2006size, das2016microstructure} at small deformation. 

The $\mum{1}$ domain size for the plastically constrained case displays the hardest response. This is because the constrained boundary conditions lead to gradients in the plastic strain rate, $\bfL^p$, near the boundaries, as explained below, and these gradients are larger for the smaller domain sizes (by simple scaling arguments). Of course, the presence of $\bfalpha$ also gives rise to additional plastic strain rate of the form $\bfalpha \times \bfV$ in MFDM which is a softening effect, but the net effect is one of hardening in overall response.

We now explain the reason for the development of inhomogeneity in the $\bfalpha$ field with the onset of plasticity for the case of constrained boundary conditions. This emergence of inhomogeneity can be attributed to the fact that the no-flow boundary condition induces gradients in $\bfL^p$ which lead to the evolution of $\bfalpha$ in the domain. For example, taking $\bfn = \bfe_2$ on the top boundary with $\bfalpha = \bfzero$ instantaneously, a no-flow boundary implies 
\begin{align}
&(\bfL^p \times \bfn) = \bf0 \nonumber \text{ on } ~\partial \varOmega \nonumber \\
\implies & \begin{bmatrix}
0 & 0 & L^p_{11}\\
0 & 0 & L^p_{21}\\
L^p_{33} & 0 & 0
\end{bmatrix} = \bf0 
\label{eq:LpXn_top}
\end{align}
on the top/bottom boundary whereas there is  no such constraint on $\bfL^p$ in the interior of the domain. This induces a gradient in the $L^p_{21}$  and  $L^p_{11}$ components of the plastic strain rate $\bfL^p$ in the $x_2$ direction near the top and bottom boundaries, which contributes to the development of $\alpha_{23}$ and $\alpha_{13}$ in the domain. On the left and right boundaries similar considerations hold, but weakened with the progress of deformation, as can be seen below. The normal, $\bfn = (n_1, n_2)$, changes direction with deformation and
\begin{align}
&(\bfL^p \times \bfn) = \bf0 \text{ on } ~\partial \varOmega \nonumber \\
\implies & \begin{bmatrix}
0 & 0 & L^p_{11} n_2 - L^p_{12} n_1\\
0 & 0 & L^p_{21} n_2 - L^p_{22} n_1\\
L^p_{33} n_2 & L^p_{33} n_1 & 0
\end{bmatrix} = \bf0
\label{eq:LpXn_left}
\end{align}
which implies that at small deformation ($n_1 = \pm1, n_2 = 0$), $L^p_{12} $ is constrained at the boundary. This gives rise to a gradient of $L^p_{12}$ in the $x_1$ direction which contributes to the development of $\alpha_{13}$.  As the normal changes direction, the linear constraints $L^p_{11}n_2 - L^p_{12} n_1 = 0$ and $L^p_{21} n_2 - L^p_{22} n_1 = 0$ have to hold which allow more freedom in accommodating deformation but,  nevertheless, gradients do develop.

\begin{figure}[htbp]
\centering
\includegraphics[width=.495\linewidth]{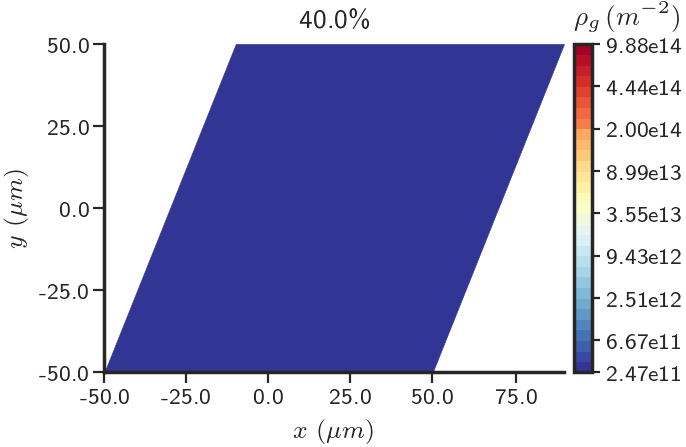}
\caption{$\rho_g$ for the $\mum{100}$ domain size with plastically unconstrained boundaries at $40\%$ strain and $\theta_0 = \mdeg{30}$ ($n_{sl} = 3$).}
    \label{fig:p1_100mc_uc_3slip_30_an_40pcnt}
 \end{figure}

Before moving on to presenting the results for the emergence of dislocation patterns in the presence of external loads, we verify that the solution for the larger domain size is close to the one obtained from conventional plasticity theories. For the larger domain size of $\mum{100}$ with unconstrained boundaries, the dislocation density norm at $40 \%$ strain is shown in figure \ref{fig:p1_100mc_uc_3slip_30_an_40pcnt}. As can be seen, the deformation is homogeneous, which is similar to the  prediction of conventional plasticity theories. The stress strain curve therefore also overlaps the conventional plasticity result as shown in Fig.~\ref{fig:se_ss}.

\subsubsection{Dislocation microstructure}
\begin{figure}[htbp]
    \centering
    \begin{subfigure}[b]{.495\linewidth}
        \centering
{\includegraphics[width=0.9\linewidth]{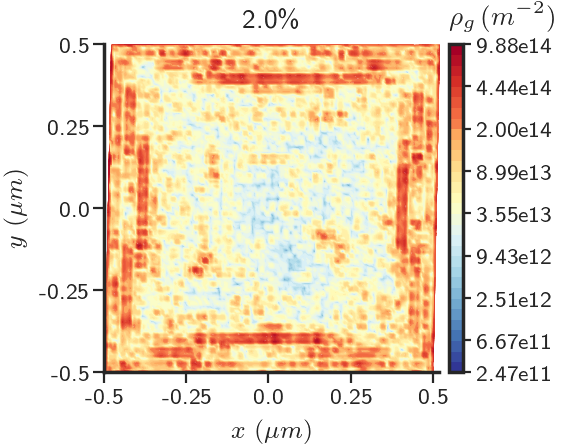}}
    \end{subfigure}%
    \begin{subfigure}[b]{.495\linewidth}
        \centering
{\includegraphics[width=0.9\linewidth]{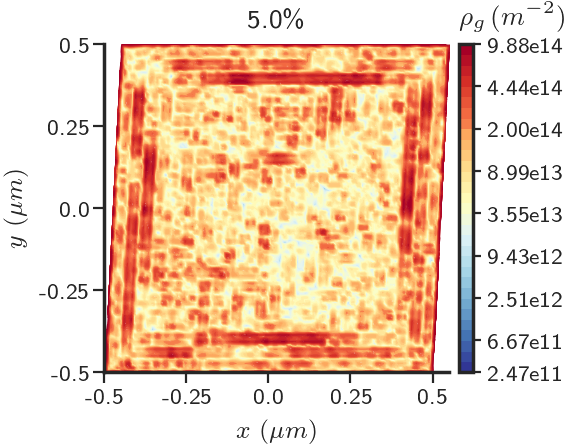}}
    \end{subfigure}\\
    \begin{subfigure}[b]{.495\linewidth}
        \centering
{\includegraphics[width=0.9\linewidth]{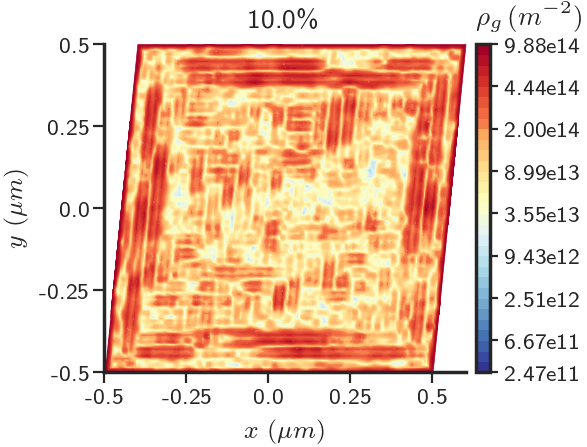}}
    \end{subfigure}%
    \begin{subfigure}[b]{.495\linewidth}
        \centering
{\includegraphics[width=0.9\linewidth]{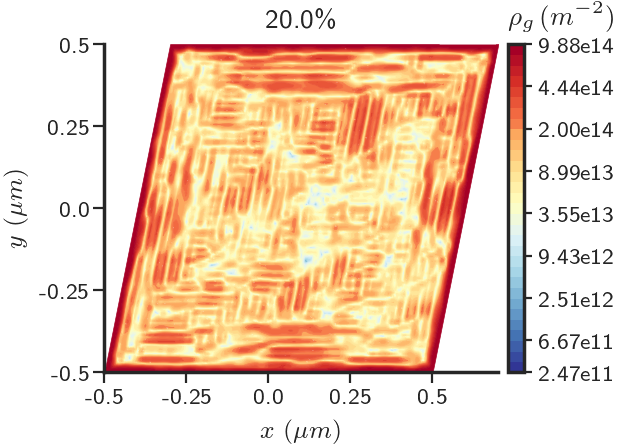}}
    \end{subfigure}\\
    \begin{subfigure}[b]{.495\linewidth}
        \centering
{\includegraphics[width=0.9\linewidth]{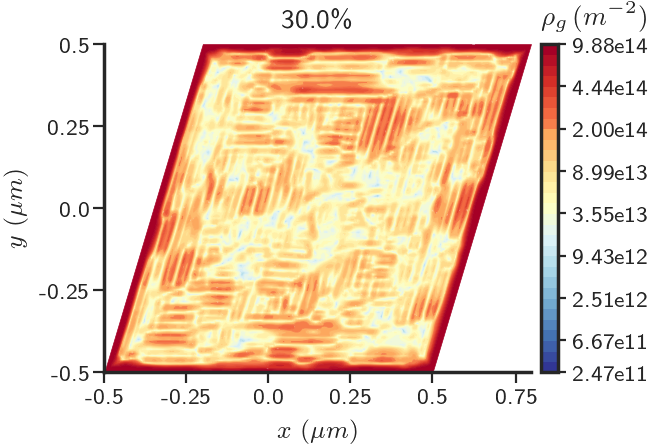}}
    \end{subfigure}%
    \begin{subfigure}[b]{.495\linewidth}
        \centering
{\includegraphics[width=0.9\linewidth]{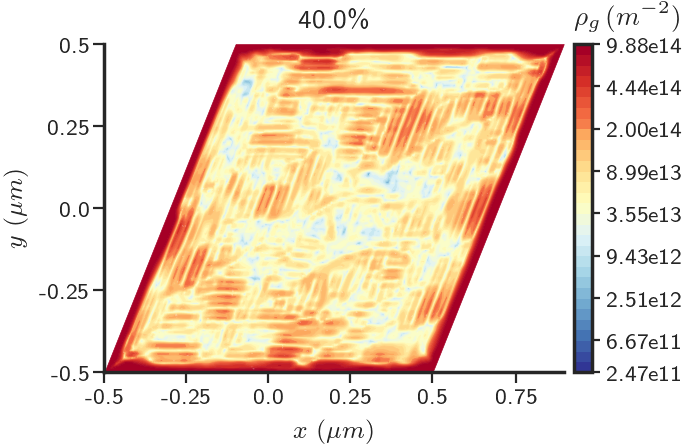}}
    \end{subfigure}\\
        \begin{subfigure}[b]{.495\linewidth}
        \centering
{\includegraphics[width=0.9\linewidth]{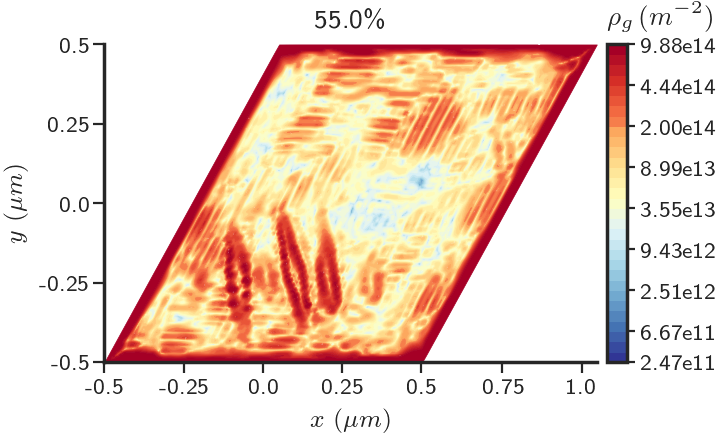}}
    \end{subfigure}%
    \begin{subfigure}[b]{.495\linewidth}
        \centering
{\includegraphics[width=0.9\linewidth]{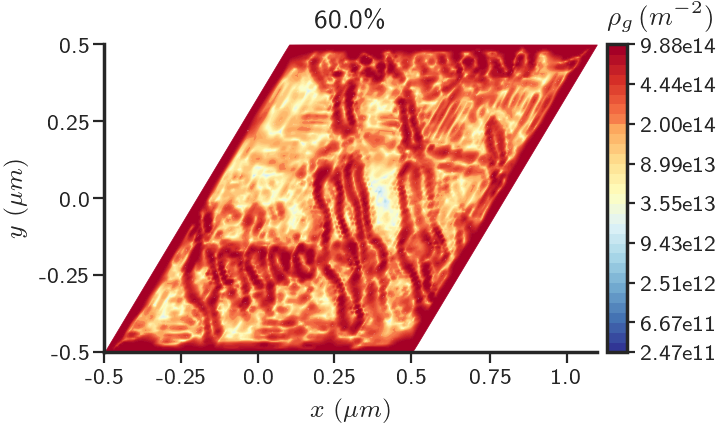}}
    \end{subfigure}%
    \caption{$\rho_g$  for the  $\mum{1}$ domain size at different strains with plastically constrained boundaries and $\theta_0 = \mdeg{30}$ ($n_{sl} = 3$).}
            \label{fig:p1_1mc_c_3slip_30_an}
 \end{figure}

\begin{figure}[htbp]
    \centering
    \begin{subfigure}[b]{.495\linewidth}
        \centering
{\includegraphics[width=0.9\linewidth]{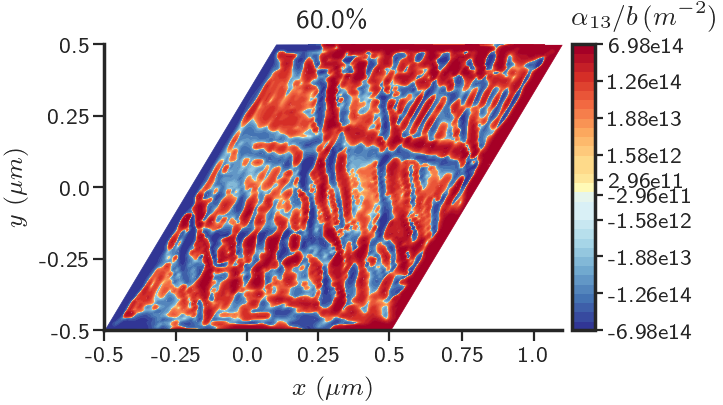}}
\caption{}
\label{fig:p1_1mc_c_3slip_30_a13}
    \end{subfigure}%
    \begin{subfigure}[b]{.495\linewidth}
        \centering
{\includegraphics[width=0.9\linewidth]{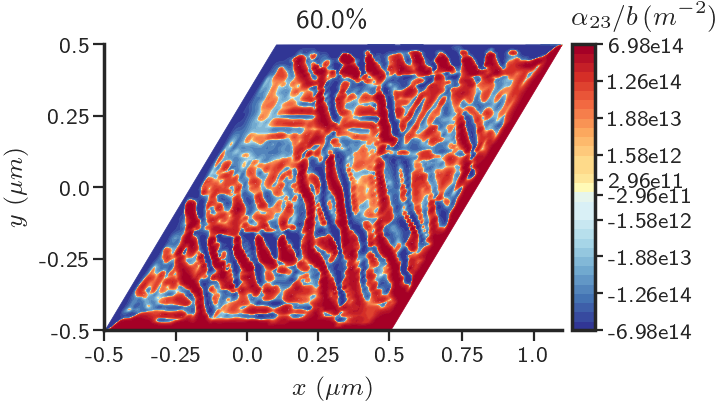}}
\caption{}
\label{fig:p1_1mc_c_3slip_30_a23}
    \end{subfigure}%
    \caption{a) $\alpha_{13}$ b) $\alpha_{23}$  for the $\mum{1}$ domain size at $60\%$ strain with plastically constrained boundaries and $\theta_0 = \mdeg{30}$  ($n_{sl} = 3$). The corresponding $\rho_g$ is shown in Fig.~\ref{fig:p1_1mc_c_3slip_30_an}.}
            \label{fig:p1_1mc_c_3slip_30_a_components}
 \end{figure}

We now present results of stressed dislocation patterns in crystal plasticity at finite deformation using MFDM. Fig.~\ref{fig:p1_1mc_c_3slip_30_an} shows the norm of the dislocation density, $\rho_g$, at various strains for the $\mum{1}$ domain size for plastically constrained boundaries.  It can be observed that
\begin{itemize}
\item Microstructural patterns start developing even before $2\%$ strain for the $\mum{1}$ domain size.
\item The dislocation density magnitude increases in the domain up to $10\%$ strain, . However, $\rho_g$ diminishes in the interior with increasing strain and becomes quite small at $40\%$ strain. 

\item At $60\%$ strain, the sample develops two prominent (dipolar) dislocation walls enclosing a distinct region of low dislocation density (by nearly two orders of magnitude), forming a dislocation cell-like structure. The dipolar nature of the walls is confirmed by looking at the magnitude of the individual dislocation components, $\alpha_{13}$ and $\alpha_{23}$ as shown in Figures \ref{fig:p1_1mc_c_3slip_30_a13} and \ref{fig:p1_1mc_c_3slip_30_a23} respectively.

\item The Burgers vector, $\bfb$, content of any area patch $A$ is given by 
\begin{align*}
\bfb = \int_A \bfalpha \bfn \, dA.
\end{align*}
An important point to note here is that in the case of plastically constrained boundaries, there is no flux of ED or SD  from  the boundary into the domain. Therefore, in the absence of any inflow or outflow flux of dislocations, and the $\bfalpha$ evolution being a conservation law \eqref{eq:mfdm_alpha} for Burgers vector, the total Burgers vector content of the whole body has to remain constant in time. Since, the initial Burgers vector content was $\bf0$ (dislocation free at $t=0$), the dislocation microstructure needs to be such that the Burgers vector (for the whole domain) remains $\bf0$ at all times, which makes the appearance of a distribution with opposite signs inevitable as shown in Figure \ref{fig:p1_1mc_c_3slip_30_a_components}. Of course, that dipolar walls should be produced is not a consequence of the conservation law and a somewhat realistic outcome of our model.
\end{itemize}

\begin{figure}[htbp]
    \centering
    \begin{subfigure}[b]{.495\linewidth}
        \centering
{\includegraphics[width=0.9\linewidth]{./figures/SE/3slipsystem/case1/C/1mic/AN-40pcnt.png}}
\caption{}
    \end{subfigure}%
    \begin{subfigure}[b]{.495\linewidth}
        \centering
{\includegraphics[width=0.9\linewidth]{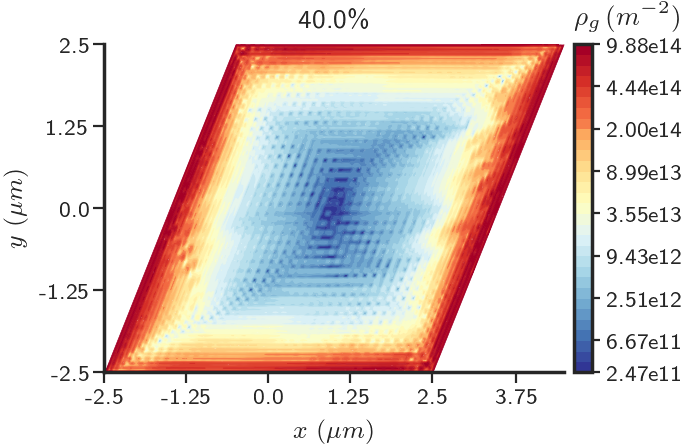}}
\caption{}
            \label{fig:p1_5mc_c_3slip_30_an_40pcnt}
    \end{subfigure}
    \caption{Comparison of $\rho_g$   for the a) $\mum{1}$ b) $\mum{5}$ sample sizes with plastically constrained boundaries at $40\%$ strain and $\theta_0 = \mdeg{30}$ ($n_{sl} = 3$).}
            \label{fig:p1_1mc_c_3slip_30_an_5mc}
 \end{figure}

The microstructural patterns for the domain size of $\mum{5}$ are significantly different from those of $\mum{1}$ domain size. Figure \ref{fig:p1_1mc_c_3slip_30_an_5mc} shows the comparison of the microstructure obtained for the $\mum{1}$ and $\mum{5}$ domain sizes at $40\%$ strain. The dislocation density is generated because of the constrained boundary conditions (as explained earlier in the discussion surrounding \eqref{eq:LpXn_top} and \eqref{eq:LpXn_left}) for both the domain sizes, but for the $\mum{5}$ domain the accumulation occurs only near the boundary. The difference can be understood by noting that the sum of the widths of the two boundary layers in the $\mum{5}$ domain add up to almost the entire linear dimension of the $\mum{1}$ domain. Assuming the patterns have an intrinsic length scale in the submicron range, as substantiated by the $\mum{1}$ results, dislocation patterns are likely to occur within the boundary layers of the $\mum{5}$ domain.

\begin{figure}[htbp]
    \centering
    \begin{subfigure}[b]{.495\linewidth}
        \centering
{\includegraphics[width=0.9\linewidth]{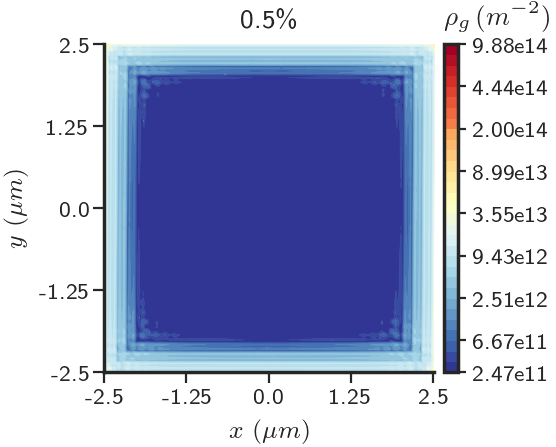}}
\caption{}
    \end{subfigure}%
    \begin{subfigure}[b]{.495\linewidth}
        \centering
{\includegraphics[width=0.9\linewidth]{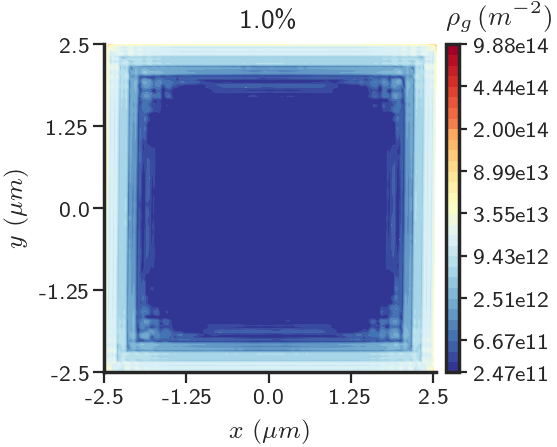}}
\caption{}
    \end{subfigure}
    \caption{ $\rho_g$ distribution for the  $\mum{5}$ domain size with plastically constrained boundaries at a) $0.5\%$ b) $1\%$ strain and $\theta_0 = \mdeg{30}$ ($n_{sl} = 3$).}
            \label{fig:p1_5mc_c_3slip_30_an_2strains}
 \end{figure}

At finite strains, the accumulation of dislocations develops an asymmetry along the boundaries for the $\mum{5}$ domain size. This is because of the change in orientation of the boundary normal with deformation at the left and right boundaries. This is corroborated by Figure \ref{fig:p1_5mc_c_3slip_30_an_2strains} where the dislocation density distribution is symmetric at small applied strain (the asymmetry in Fig.~\ref{fig:p1_5mc_c_3slip_30_an_2strains} persists at large applied strain).
 
\subsubsection{\texorpdfstring{$ssd$}{ssd} distribution}
\begin{figure}[htbp]
    \centering
    \begin{subfigure}[b]{.495\linewidth}
        \centering
{\includegraphics[width=0.9\linewidth]{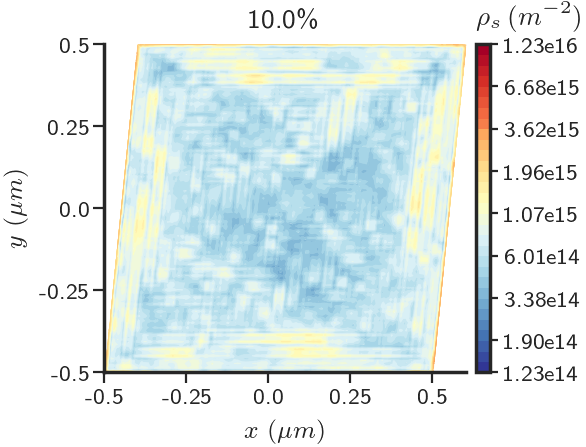}}
    \end{subfigure}%
    \begin{subfigure}[b]{.495\linewidth}
        \centering
{\includegraphics[width=0.9\linewidth]{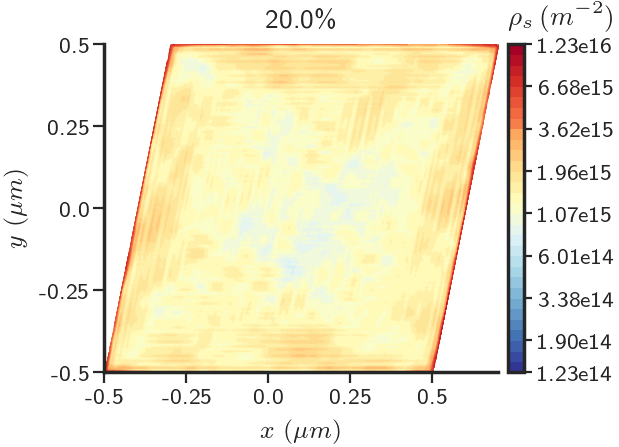}}
    \end{subfigure}\\
    \begin{subfigure}[b]{.495\linewidth}
        \centering
{\includegraphics[width=0.9\linewidth]{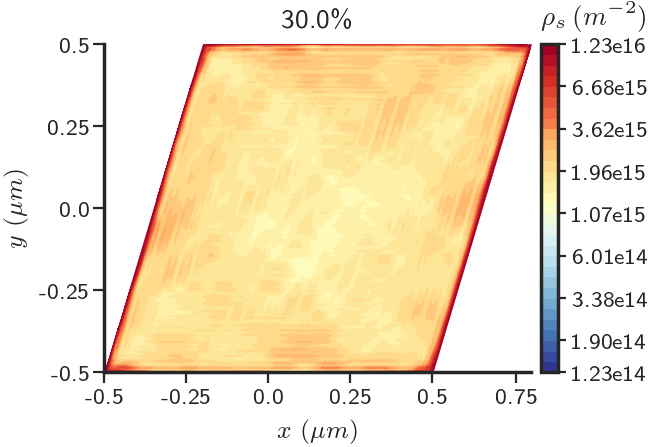}}
    \end{subfigure}%
    \begin{subfigure}[b]{.495\linewidth}
        \centering
{\includegraphics[width=0.9\linewidth]{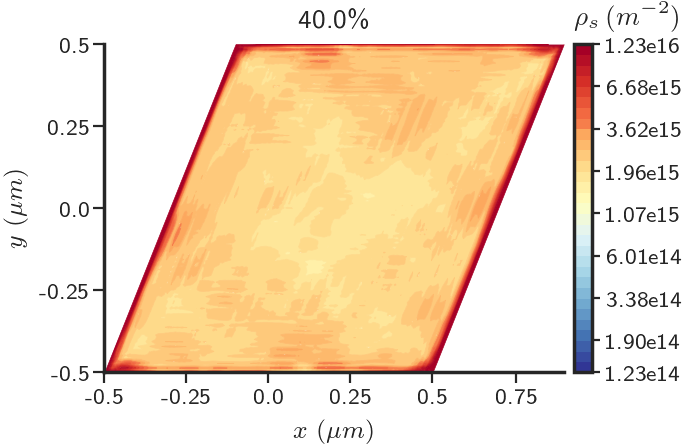}}
    \end{subfigure}\\
    \begin{subfigure}[b]{.495\linewidth}
        \centering
{\includegraphics[width=0.9\linewidth]{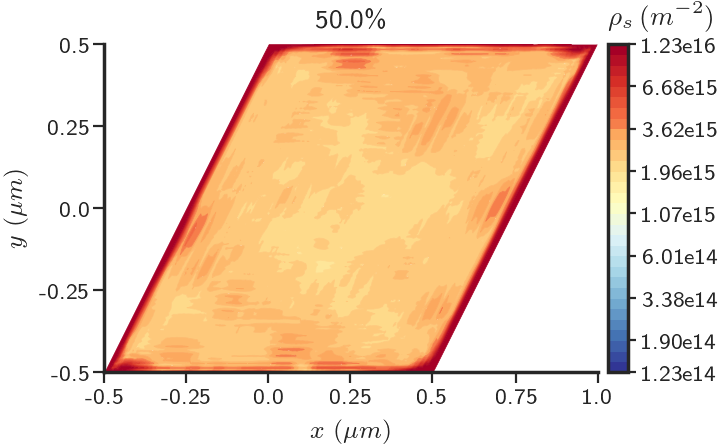}}
    \end{subfigure}%
    \begin{subfigure}[b]{.495\linewidth}
        \centering
{\includegraphics[width=0.9\linewidth]{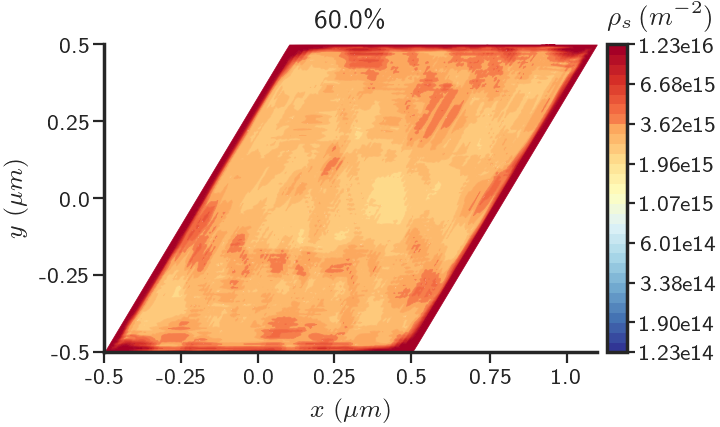}}
    \end{subfigure}%
        \caption{$ssd$ distribution  for the $\mum{1}$ domain size at different strains with plastically constrained boundaries and $\theta_0 = \mdeg{30}$ ($n_{sl} = 3$).}
            \label{fig:p1_1mc_c_3slip_30_ssd}
 \end{figure}

A large part of the plastic strain rate at mesoscales  comes from expansion of unresolved dislocation loops that constitute SD. The $ssd$, $\rho_s$, given by Eq.~\eqref{eq:ssd} in MFDM, is proportional to the root-mean-square of the SD. Figure \ref{fig:p1_1mc_c_3slip_30_ssd} presents the distribution of $ssd$ in the domain for the  $\mum{1}$ sample size with plastically constrained boundaries. The distribution is mildly patterned in the domain, with magnitude increasing with strain. There is a variation of at least an order of magnitude in the domain at all strain levels.

\subsubsection{Unloaded stressed microstructures}
\label{sec:unloaded}
The $\mum{1}$ domain is unloaded from $60\%$ strain by reversing the loading direction. The boundary conditions for velocity are taken as $v_2 = 0$ and $v_1 = -\hat\mGamma y$ until $\tau$ (averaged stress on the top surface) becomes zero. Then, we decrease the nodal reaction forces steadily over time until a tolerance of $\max_{j} \{\text{abs} (F_j)\} < 10^{-4} \times (g_0\,h) $ is reached, where $\{\text{abs}(F_j)\}$  refers to the absolute value of the $j^{th}$ entry in the nodal reaction force array $\{F\}$, defined in Sec.~\ref{sec:algorithm}, with size equal to number of degrees of freedom where (material) velocity Dirichlet-boundary conditions are applied, and $h$ is the element size. Thereafter, we let the system achieve thermodynamic equilibrium by requiring all evolution, i.e., of $\bfalpha$, $\bff$, $g$, to become small. Hence, if at all, these microstructures evolve very slowly. 

\begin{figure}
       \begin{subfigure}[b]{.495\linewidth}
        \centering
{\includegraphics[width=0.9\linewidth]{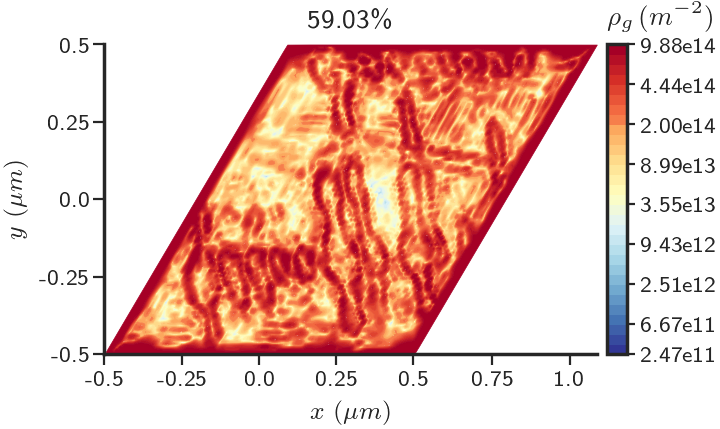}}
\caption{}
            \label{fig:p1_unloaded_an}
    \end{subfigure}%
    \begin{subfigure}[b]{.495\linewidth}
        \centering
{\includegraphics[width=0.9\linewidth]{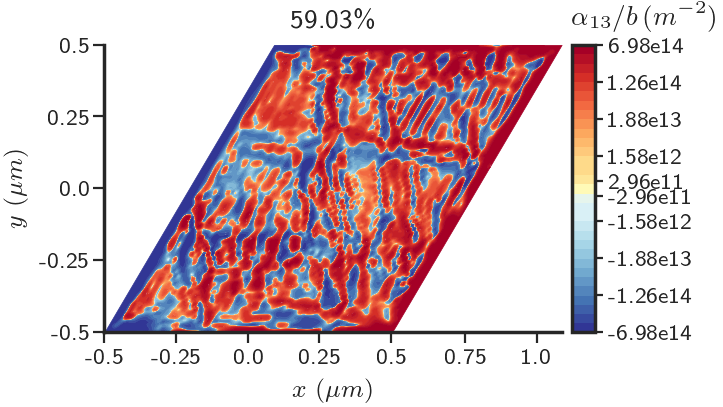}}
\caption{}
\label{fig:p1_unloaded_a13}
    \end{subfigure}\\
        \begin{subfigure}[b]{.99\linewidth}
        \centering
{\includegraphics[width=0.5\linewidth]{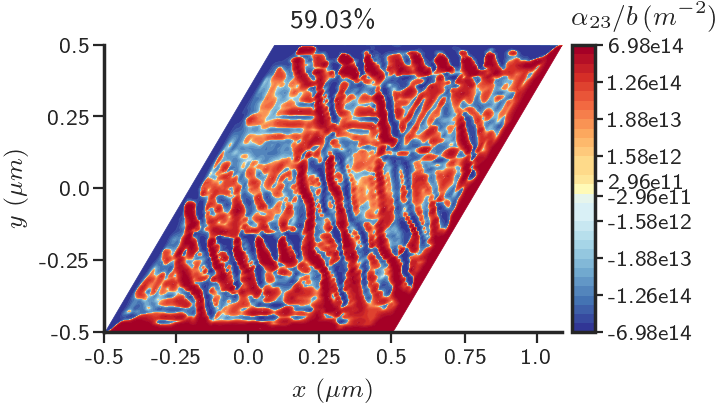}}
\caption{}
\label{fig:p1_unloaded_a23}
    \end{subfigure}%
    \caption{Unloaded, stressed microstructure a) $\rho_g$ b) $\alpha_{13}$ c) $\alpha_{23}$ for the  $\mum{1}$ domain size with plastically constrained boundaries and $\theta_0 = \mdeg{30}\ (n_{sl} = 3)$.}
            \label{fig:p1_unloaded}
 \end{figure}

The $\rho_g$ distribution on the unloaded configuration is shown in Fig.~\ref{fig:p1_unloaded_an}. The $\alpha_{13}$ and $\alpha_{23}$ components of the dislocation density tensor are shown in Figures \ref{fig:p1_unloaded_a13} and \ref{fig:p1_unloaded_a23}, respectively.

\begin{figure}
    \begin{subfigure}[b]{.495\linewidth}
        \centering
{\includegraphics[width=0.9\linewidth]{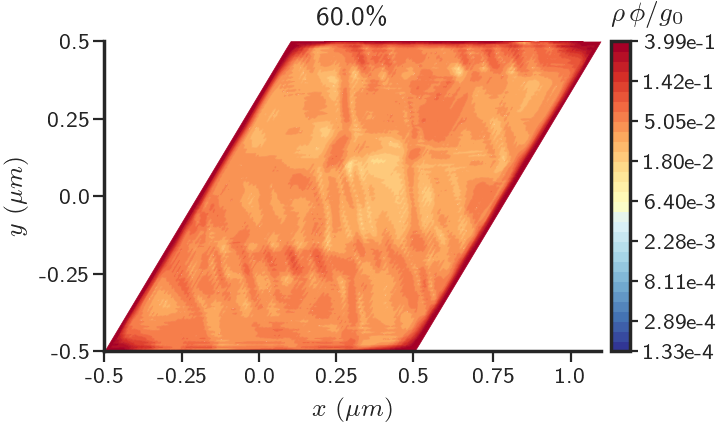}}
\caption{}
\label{fig:p1_unloaded_a13_SE}
    \end{subfigure}
        \begin{subfigure}[b]{.495\linewidth}
        \centering
{\includegraphics[width=0.9\linewidth]{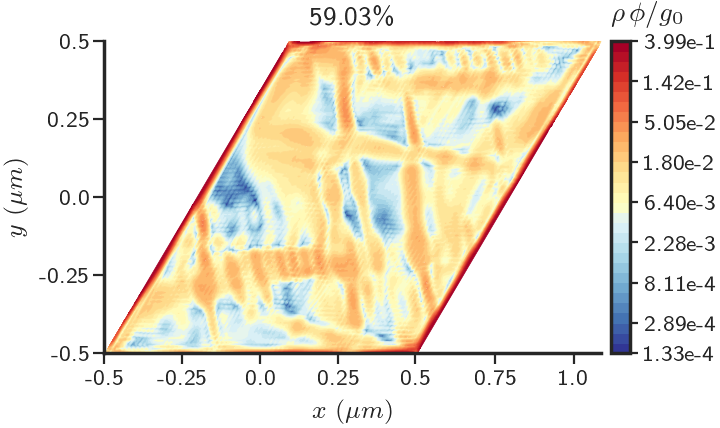}}
\caption{}
\label{fig:p1_unloaded_a23_SE}
    \end{subfigure}%
    \caption{Non-dimensional elastic energy density $\frac{\rho \phi}{g_0}$ for the unloaded $\mum{1}$ domain size with plastically constrained boundaries and $\theta_0 = \mdeg{30}\ (n_{sl} = 3)$ a) before  b) after unloading.}
            \label{fig:p1_Energy}
 \end{figure}

In Fig.~\ref{fig:p1_Energy}, we plot the elastic energy density in the domain, given by $\rho \phi$ \eqref{eq:phi_W}, before and after the unloaded equilibration. The energy density variation in the body after unloaded equilibration is at least an order of magnitude smaller in most of the interior of the domain than the energy density before unloading.

\begin{figure}
    \begin{subfigure}[b]{.495\linewidth}
        \centering
{\includegraphics[width=0.9\linewidth]{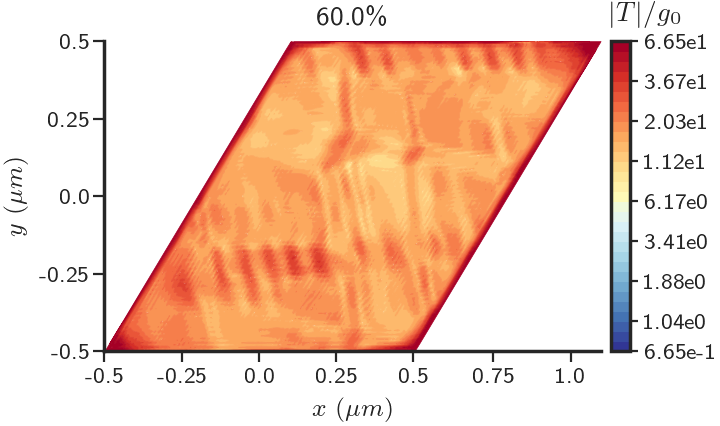}}
\caption{}
\label{fig:p1_unloaded_a13_Tnorm}
    \end{subfigure}
        \begin{subfigure}[b]{.495\linewidth}
        \centering
{\includegraphics[width=0.9\linewidth]{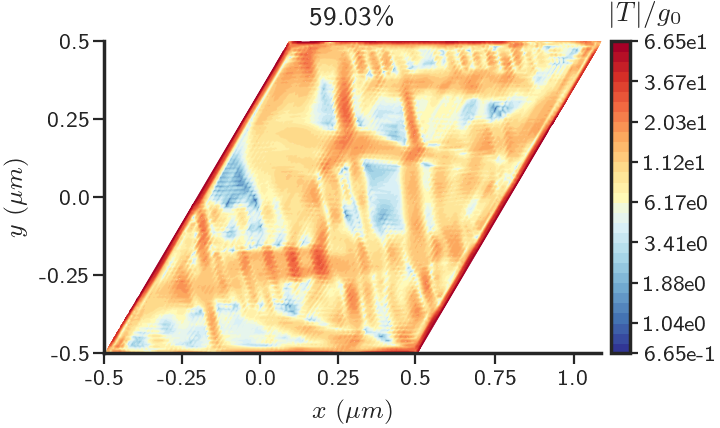}}
\caption{}
\label{fig:p1_unloaded_a23_Tnorm}
    \end{subfigure}%
    \caption{Non-dimensional stress norm $\frac{|\bfT|}{g_0}$ for the $\mum{1}$ domain size with plastically constrained boundaries and $\theta_0 = \mdeg{30}\ (n_{sl} = 3)$ a) before b) after unloading.}
            \label{fig:p1_Tnorm}
 \end{figure}

Figure \ref{fig:p1_Tnorm} shows the non-dimensionalized norm of the stress field in the domain before and after unloaded equilibration. It can be seen that the body is not stress-free after equilibration, and is stressed upto $\approx 6$ times the initial yield strength $g_0$. The corresponding plastic strain rate magnitudes, $\hat\gamma^k$ for $k^{th}$ slip system, are found to be negligible,  $\frac{\hat{\gamma}^k}{\hat{\gamma}_0} \approx 10^{-8}$.

Hence, we conclude that the unloaded stressed microstructures are kinetically trapped, (computational) metastable equilibrium solutions of the theory.

We remark that \emph{the entire class of ED distributions arising from spatially heterogeneous rotation distributions in unloaded bodies constitute exact equilibria of our model}, since they result in vanishing stress fields. This results in vanishing $\hat{\gamma}^k$ on any slip system which implies $\bfV = \bfzero$ and $\bfL^p= \bfzero$ from \eqref{eq:V_zeta} and \eqref{eq:Lp}, respectively, and consequently $\bfalpha, \bff, g$ cease to evolve from such states.

\subsubsection{Microstructure with unconstrained boundary conditions}\label{sec:unconstrained}
\begin{figure}[htbp]
        \centering
        \begin{subfigure}[b]{.495\linewidth}
        \centering
{\includegraphics[width=0.9\linewidth]{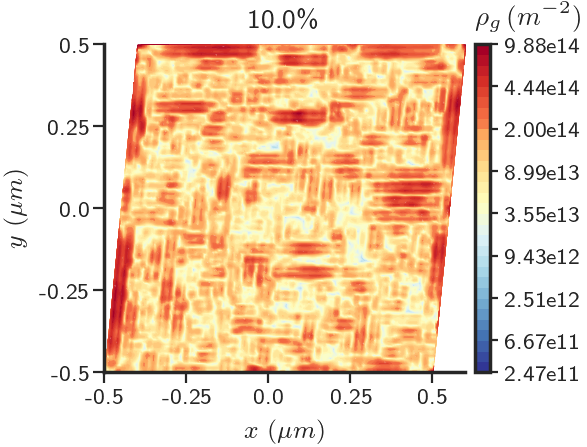}}
    \end{subfigure}
        \begin{subfigure}[b]{.495\linewidth}
        \centering
{\includegraphics[width=0.9\linewidth]{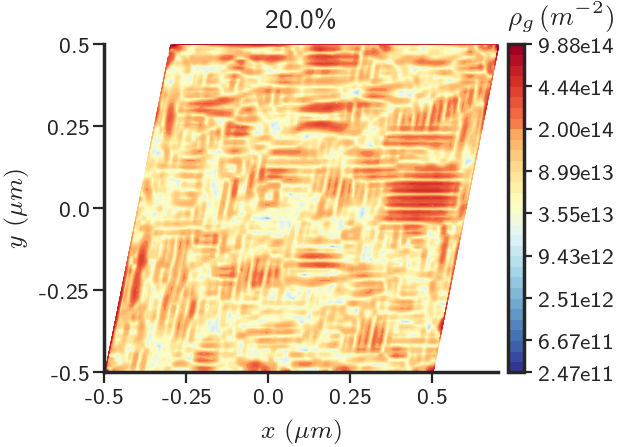}}
    \end{subfigure}\\
    \begin{subfigure}[b]{.495\linewidth}
        \centering
{\includegraphics[width=0.9\linewidth]{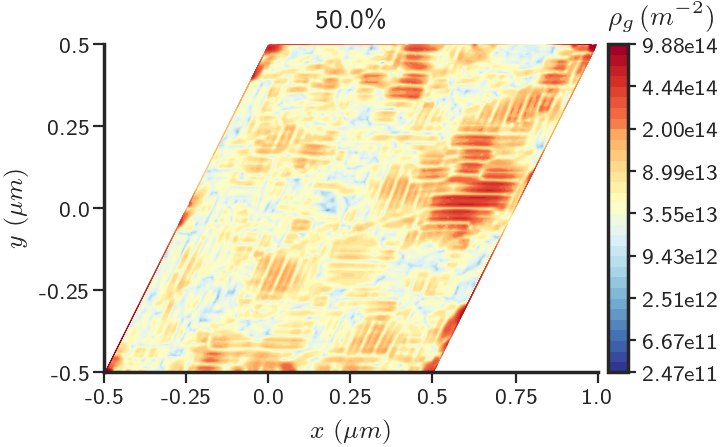}}
    \end{subfigure}
        \begin{subfigure}[b]{.495\linewidth}
        \centering
{\includegraphics[width=0.9\linewidth]{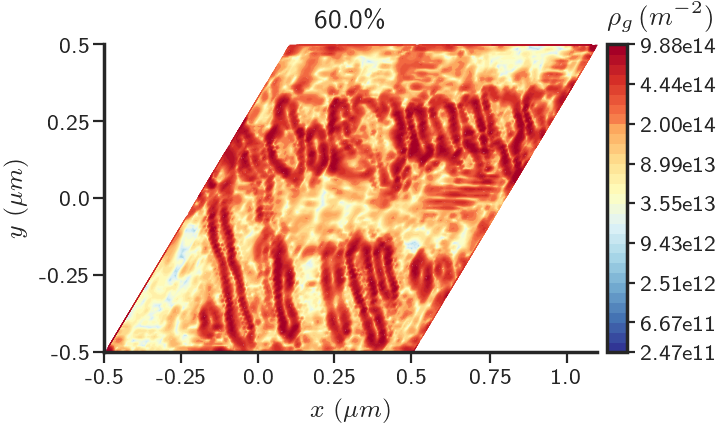}}
    \end{subfigure}%
    \caption{ $\rho_g$ at different strains for the $\mum{1}$ domain size with plastically unconstrained boundaries and $\theta_0 = \mdeg{30}\ (n_{sl} = 3)$.}
            \label{fig:p1_1mc_uc_3slip_30_an}
 \end{figure}

We demonstrate that the emergence of patterning for the $\mum{1}$ domain size is not dependent on the condition that the boundaries be plastically constrained. Figure \ref{fig:p1_1mc_uc_3slip_30_an} shows the dislocation pattern in the $\mum{1}$ domain size at different strains with unconstrained boundaries  ($\theta_0 = \mdeg{30}, n_{sl} = 3$). After an initial burst at relatively small strains, the patterns again become pronounced at $60\%$ strain as was the case for constrained boundaries presented Fig.~\ref{fig:p1_1mc_c_3slip_30_an}.

\subsubsection{Effect of slip system orientation on microstructure and stress response}
\label{sec:orientation_45_n3}
We now explore the question of variation in the microstructural patterns when the initial lattice orientation, $\theta_0$,  is changed. Keeping all parameters as in Table \ref{tab:se_parameters} except for setting $\theta_0$ to $\mdeg{45}$, we obtain  microstructural patterns for the $\mum{1}$ domain size shown below in Fig.~\ref{fig:p1_1mc_c_3slip_45_an} that are very similar to the microstructure in Fig.~\ref{fig:p1_1mc_c_3slip_30_an} obtained for $\theta_0 = \mdeg{30}$. This is because the applied averaged simple shear deformation can be accommodated by three independent slip systems regardless of orientation (in fact two suffices for incompressible velocity fields).

\begin{figure}[htbp]
    \centering
    \begin{subfigure}[b]{.495\linewidth}
        \centering
{\includegraphics[width=0.9\linewidth]{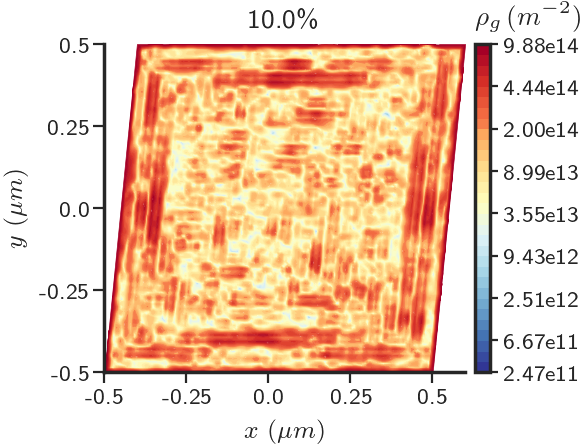}}
    \end{subfigure}%
    \begin{subfigure}[b]{.495\linewidth}
        \centering
{\includegraphics[width=0.9\linewidth]{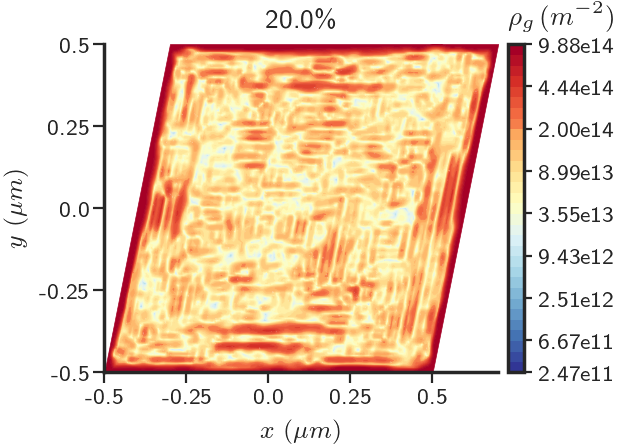}}
    \end{subfigure}\\
    \begin{subfigure}[b]{.495\linewidth}
        \centering
{\includegraphics[width=0.9\linewidth]{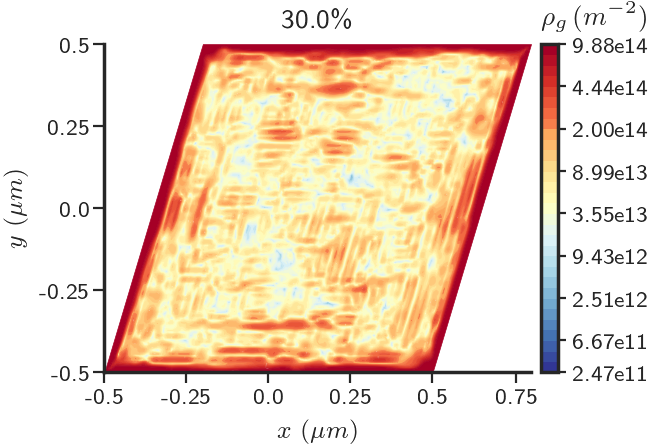}}
    \end{subfigure}%
    \begin{subfigure}[b]{.495\linewidth}
        \centering
{\includegraphics[width=0.9\linewidth]{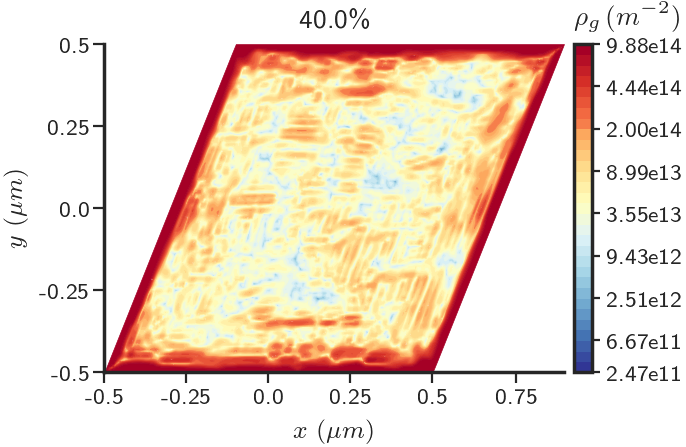}}
    \end{subfigure}\\
        \begin{subfigure}[b]{.495\linewidth}
        \centering
{\includegraphics[width=0.9\linewidth]{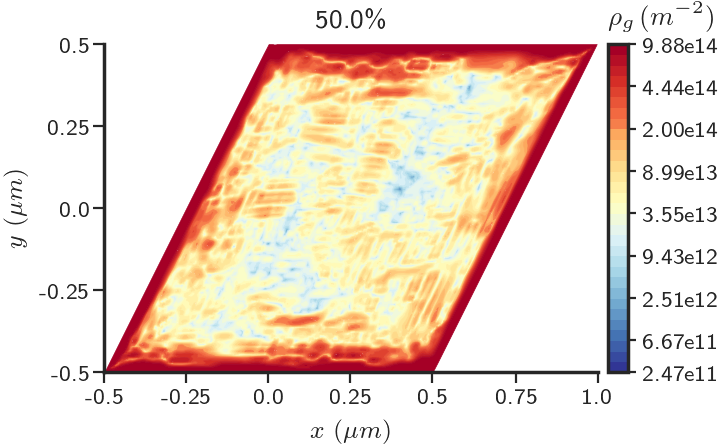}}
    \end{subfigure}%
    \begin{subfigure}[b]{.495\linewidth}
        \centering
{\includegraphics[width=0.9\linewidth]{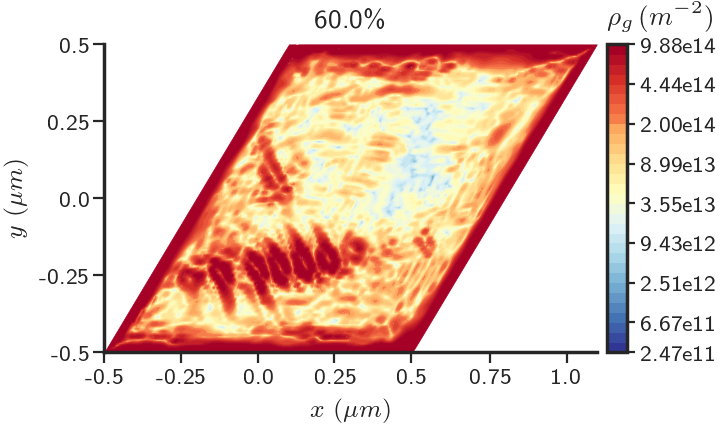}}
    \end{subfigure}%
    \caption{$\rho_g$ distribution for the  $\mum{1}$ domain size at different strains with plastically constrained boundaries and $\theta_0 = \mdeg{45}$ ($n_{sl} = 3$).}
            \label{fig:p1_1mc_c_3slip_45_an}
 \end{figure}

\begin{figure}[htbp]
        \centering
{\includegraphics[width=.7\linewidth]{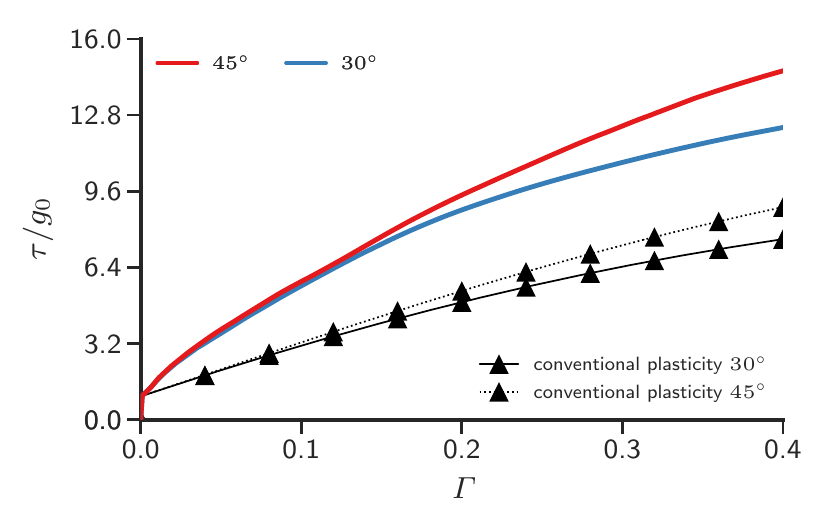}}
        \caption{Stress-strain response for the $\mum{1}$ domain size for $\theta_0 = \mdeg{30}$ and $\theta_0 = \mdeg{45}$ ($n_{sl} = 3$).}
            \label{fig:p1_se_ss_3slip_45}
    \end{figure}

Figure \ref{fig:p1_se_ss_3slip_45} presents a comparison of the stress-strain plots for the $\mum{1}$ domain size when the orientation of the slip system is changed from $\theta_0 = \mdeg{30}$ to $\theta_0 = \mdeg{45}$. This change in orientation results in a harder stress-strain response which can also be seen in corresponding responses modeled by conventional theory. 

\subsection{Dislocation microstructure in single slip}
\label{sec:single_slip}
Motivated by the approximate invariance of the microstructural patterns with respect to crystal orientation and our conjecture in Sec.~\ref{sec:orientation_45_n3} that the issue is related to the accommodation of the applied average deformation (rate) field by the plastic slip systems available, we now consider a body with only a single slip system. The hypothesis to be tested is that in this scenario the applied deformation cannot be accommodated, thus leading to higher stresses and elastic incompatibilities, the degree of which should depend on the slip system orientation with respect to the applied simple shear. By `accommodation' here we mean that the tensorial direction of the simple shearing motion defined by the applied boundary conditions  can be represented as a linear combination of the evolving slip-system dyads of the material, assuming active slip systems.

As before, the initial orientation of the slip system will be defined by $\theta_0$, which is the angle of the slip direction from the $x_1$ axis. The initial slip direction and normal for the slip system is given as
\begin{align*}
&\bfm_0^1 = (\cos(\theta_0), \sin(\theta_0))  &\bfn_0^1 = (-\sin(\theta_0), \cos(\theta_0)).
\end{align*}

 \subsubsection{Dislocation microstructure}
\begin{figure}[htbp]
    \centering
    \begin{subfigure}[b]{.495\linewidth}
        \centering
{\includegraphics[width=0.9\linewidth]{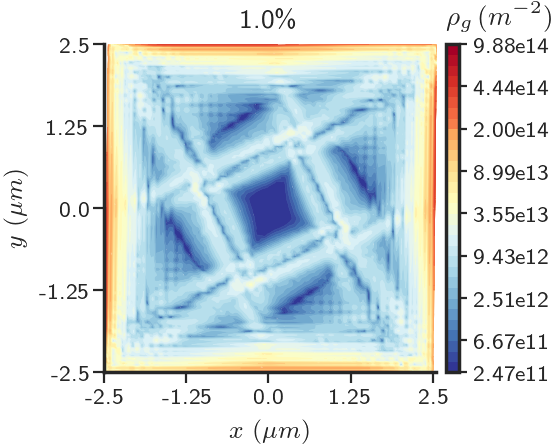}}
    \end{subfigure}%
   \begin{subfigure}[b]{.495\linewidth}
        \centering
{\includegraphics[width=0.9\linewidth]{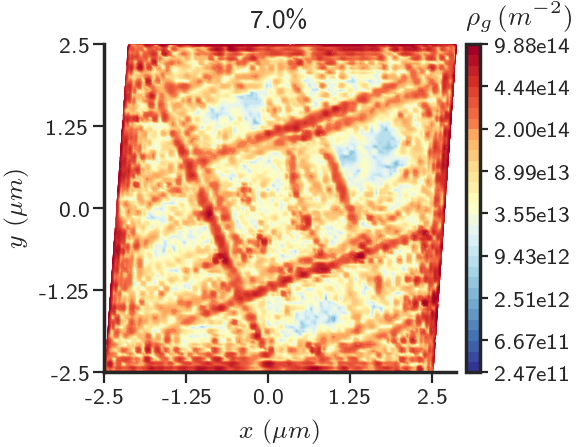}}
    \end{subfigure}
    \caption{$\rho_g$ for  the $\mum{5}$ domain size at $1\%$ and $ 7\%$ strain with plastically constrained boundaries and  $\theta_0 = \mdeg{30}$ ($n_{sl} = 1$).}
\label{fig:p1_5mc_c_1slip_30_an}
    \end{figure}

 We plot the microstructure for the $\mum{5}$ domain size at $5\%$ strain in Figure \ref{fig:p1_5mc_c_1slip_30_an} for $\theta_0 = \mdeg{30}$. We can see that for the case of a single slip system, the patterns in the $\mum{5}$ domain size are very different from those for the $3$-slip-systems case (shown in Fig.~\ref{fig:p1_5mc_c_3slip_30_an_40pcnt}). This can be attributed to the fact the the deformation now is much more constrained due to the presence of only a single slip system. In contrast to Fig.~\ref{fig:p1_5mc_c_3slip_30_an_40pcnt}, cell structures form in the interior of the domain at $5\%$ strain. Therefore, these observations substantiate our conjecture related to accommodation.

\begin{figure}[htbp]
    \centering
    \begin{subfigure}[b]{.495\linewidth}
        \centering
{\includegraphics[width=0.9\linewidth]{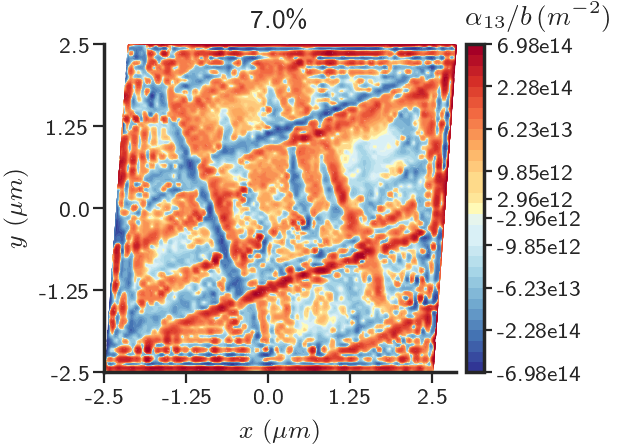}}
\caption{}
    \end{subfigure}%
   \begin{subfigure}[b]{.495\linewidth}
        \centering
{\includegraphics[width=0.9\linewidth]{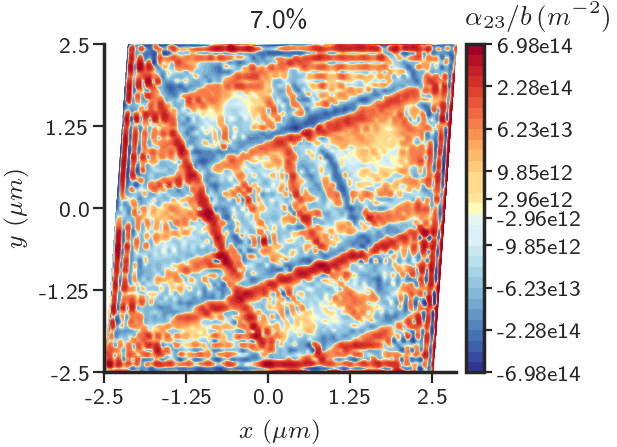}}
\caption{}
    \end{subfigure}
    \caption{a) $\alpha_{13}$ b) $\alpha_{23}$ for the  $\mum{5}$ domain size at $ 7\%$ strain with plastically constrained boundaries and  $\theta_0 = \mdeg{30}$ ($n_{sl} = 1$).}
\label{fig:p1_5mc_c_1slip_30_acomponents}
    \end{figure}

Figure \ref{fig:p1_5mc_c_1slip_30_acomponents} shows the individual components of the dislocation density tensor for $\mum{5}$ domain size at $5\%$ strain. We can notice  monopolar walls, of both types ($\alpha_{13}$ and $\alpha_{23}$) of dislocations, forming in the interior of the domain. If we relate these to the norm of the dislocation density tensor shown in Fig.~\ref{fig:p1_5mc_c_1slip_30_an}, we can see that these monopolar walls are the boundary of the cell structure formed in the center of the domain. Of note is also the dipolar wall in kink orientation \cite[Sec. IV A]{asaro1983micromechanics} to the primary (and only) slip plane.

\subsubsection{Effect of slip system orientation on microstructure and stress response}
\begin{figure}[htbp]
    \centering
    \begin{subfigure}[b]{.495\linewidth}
        \centering
{\includegraphics[width=0.9\linewidth]{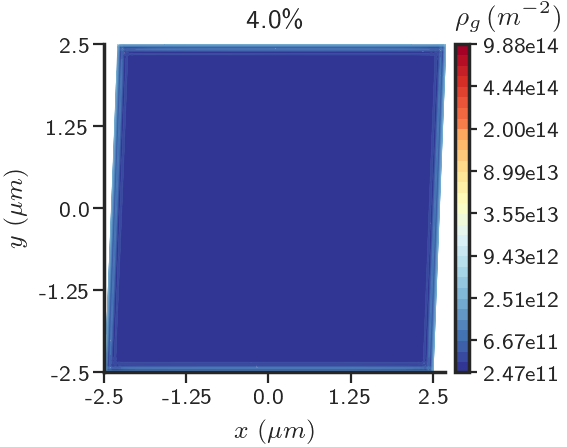}}
    \end{subfigure}%
    \begin{subfigure}[b]{.495\linewidth}
        \centering
{\includegraphics[width=0.9\linewidth]{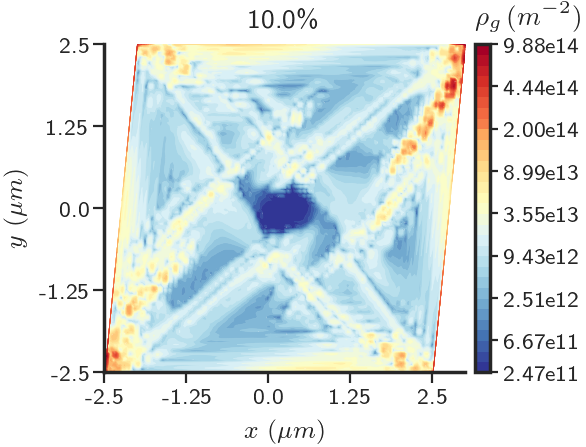}}
    \end{subfigure}
    \caption{$\rho_g$ for the $\mum{5}$ domain size at different strains with plastically constrained boundaries and $\theta_0 = \mdeg{45}$ ($n_{sl} = 1$).}
            \label{fig:p1_5mc_c_1slip_45_an}
 \end{figure}

Here we explore the change in the microstructural pattern for the $\mum{5}$ domain size when the orientation of the slip system is changed  from $\mdeg{30}$ to $\mdeg{45}$. Figure \ref{fig:p1_5mc_c_1slip_45_an} shows the obtained patterns for $\theta_0 = \mdeg{45}$. It can be seen that the $\rho_g$ becomes significant only at strains larger than $5\%$. This is because at small deformation, the resolved shear stress on the slip system, with instantaneous orientation denoted by $\theta$, is given as $\tau = \bfm\cdot\bfT\bfn = T_{12} \cos(2\theta)$ which is small for  $\theta \approx \mdeg{45}$ so that the applied deformation has to be elastically accommodated. As the deformation progresses, the lattice rotation affects the slip system orientation $\theta$. This change in orientation of the slip system produces a small plastic strain rate on it and results in the development of ED because of the plastically constrained boundary condition as explained earlier in the discussion surrounding \eqref{eq:LpXn_top} and \eqref{eq:LpXn_left}. However, even though the plastic strain gradients are large, the plastic strain itself is small enough to not cause any noticeable change from elastic behavior in the stress-strain response.

\begin{figure}[htbp]
        \centering
{\includegraphics[width=.6\linewidth]{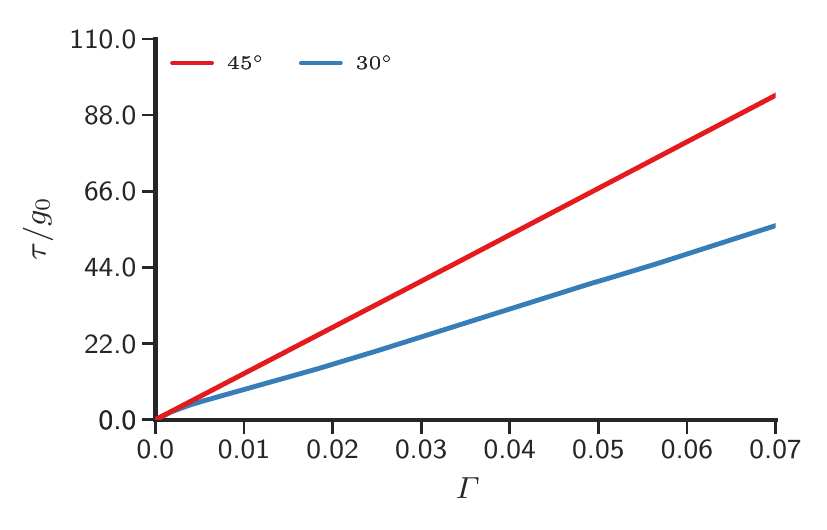}}
        \caption{Stress-strain response for the $\mum{5}$ domain size for plastically constrained boundaries and $\theta_0 = \mdeg{30}$ and $\theta_0 = \mdeg{45}$ ($n_{sl} = 1$).}
            \label{fig:se_ss_1slip_orientation}
    \end{figure}

A comparison of the stress-strain behavior for the $\mum{5}$ domain size when the initial orientation $\theta_0$ of the slip system is changed from $\mdeg{30}$ to $\mdeg{45}$ is presented in Fig.~\ref{fig:se_ss_1slip_orientation}. As can be seen, the response is almost $50$  times  harder than the corresponding data shown in Fig.~\ref{fig:se_ss}. As already explained above, this is the result of the elastic accommodation of the initial deformation. 

\subsection{Convergence}
\label{sec:convergence}
This section deals with the study of convergence in stress-strain response and microstructural patterns obtained in Sec.~\ref{sec:multiple_slip} ($n_{sl} = 3$).

\subsubsection{Stress-strain  response}
We study the convergence of the stress-strain responses for the $\mum{1}$ and $\mum{5}$  domain sizes with plastically constrained boundaries. The details of the meshes used in this section are as follows. For the $\mum{1}$ and $\mum{5}$ domain sizes, we use two uniform meshes of $70 \times 70$ and $140\times 140$ elements, referred to as the coarse and fine meshes, respectively. 

\begin{figure}[htbp]
    \centering
        \centering
{\includegraphics[width=.7\linewidth]{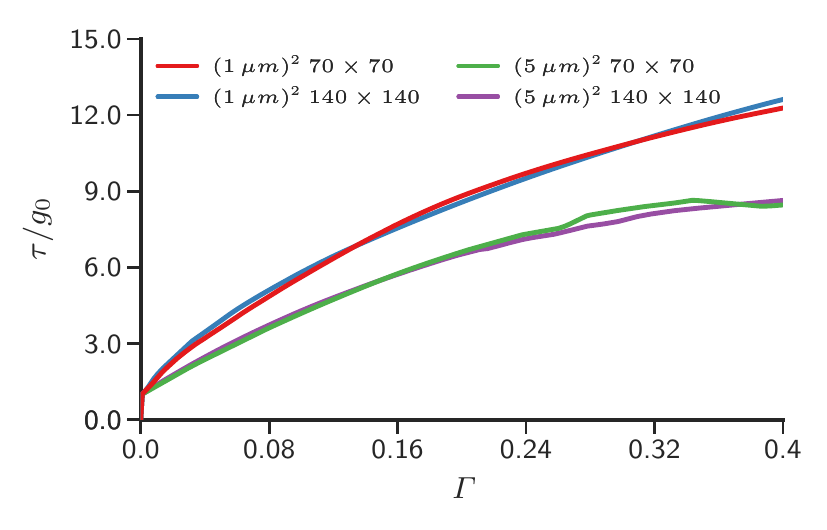}}
        \caption{Convergence in stress-strain response for the  $\mum{1}$ and $\mum{5}$ domain sizes with plastically constrained boundaries for coarse and fine meshes ($\theta_0 = \mdeg{30}, n_{sl} = 3$).}
            \label{fig:se_convergence_30}
\end{figure}

\begin{figure}
        \centering
{\includegraphics[width=.7\linewidth]{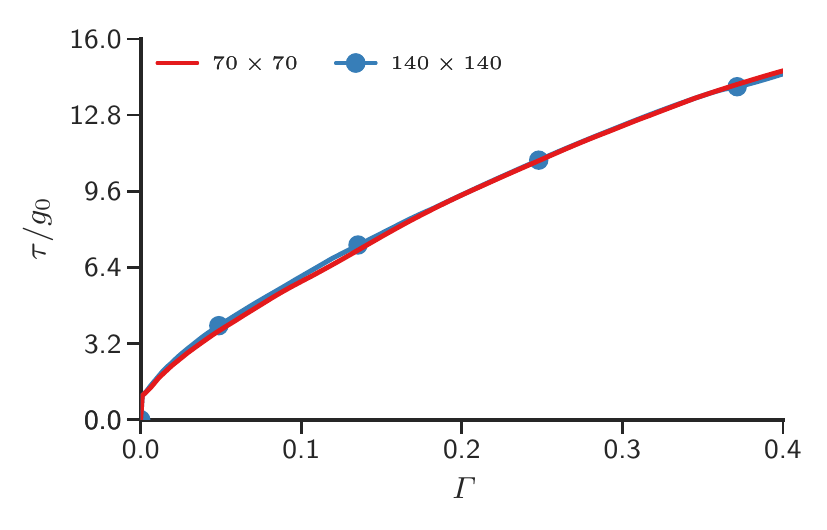}}
        \caption{Convergence in stress-strain response for the  $\mum{1}$ domain size with plastically constrained boundaries for coarse and fine meshes ($\theta_0 = \mdeg{45}, n_{sl} = 3$).}
            \label{fig:se_convergence_45}
\end{figure}

The averaged stress-strain plot for the case when the initial orientation $\theta_0$ is $\mdeg{30}$ is plotted in Fig.~\ref{fig:se_convergence_30}. The stress strain plots (almost) overlap upto $40\%$ strain for $\mum{1}$ domain size. The maximum difference (at $40\%$ strain) in the stress-strain curve is $1.2\%$, and the difference at $(28 \%)$ strain is $1.8\%$ for the smaller sample. For the case when $\theta_0 = \mdeg{45}$, the stress-strain  plots for the $\mum{1}$ sample size overlap up to $40\%$ strain and there is no discernible difference between the results obtained using coarse and fine meshes as shown in Fig.~\ref{fig:se_convergence_45}.

The unconstrained cases represent a more conservative simulation scenarios with smaller gradients than the constrained case, and hence the same mesh sizes suffice for them.

%


\subsubsection{Dislocation microstructure}
We discuss convergence of the microstructural patterns for the specific case of Sec.~\ref{sec:multiple_slip} wherein $n_{sl} = 3$ and $\theta_0 = \mdeg{30}$ (we believe that the same arguments apply to Sec.~\ref{sec:single_slip} as well).

\begin{figure}[htbp]
    \centering
    \begin{subfigure}[b]{.495\linewidth}
        \centering
{\includegraphics[width=0.9\linewidth]{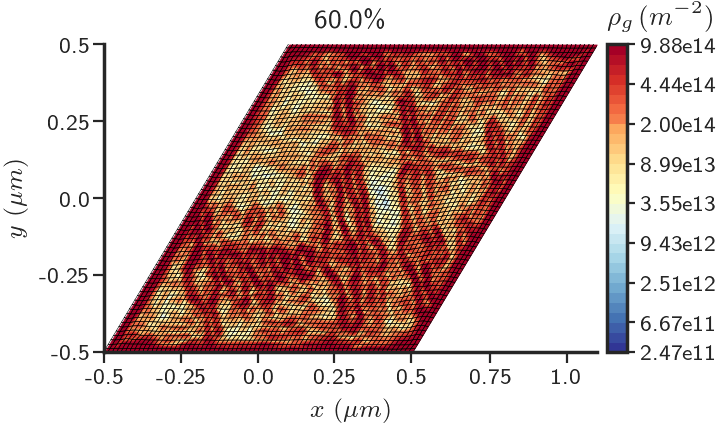}}
\caption{}
            \label{fig:p1_1mc_c_3slip_30_an_mesh}
    \end{subfigure}%
        \begin{subfigure}[b]{.495\linewidth}
        \centering
{\includegraphics[width=0.9\linewidth]{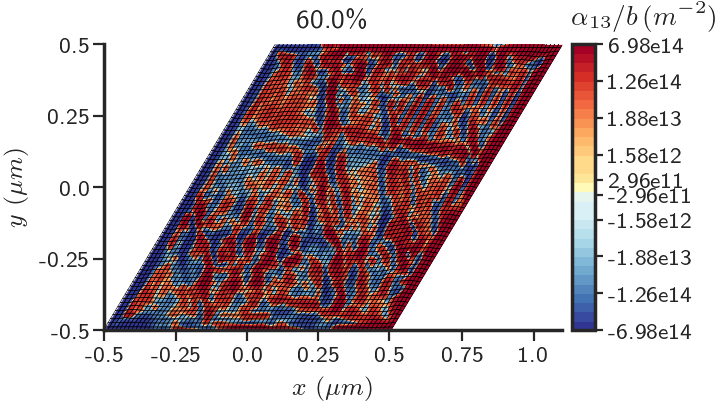}}
\caption{}
            \label{fig:p1_1mc_c_3slip_30_a13_mesh}
    \end{subfigure}\\
    \begin{subfigure}[b]{.495\linewidth}
        \centering
{\includegraphics[width=0.9\linewidth]{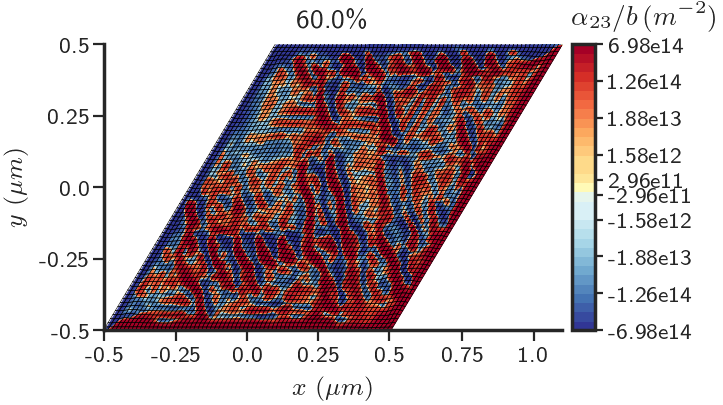}}
\caption{}
            \label{fig:p1_1mc_c_3slip_30_a23_mesh}
    \end{subfigure}%
    \begin{subfigure}[b]{.495\linewidth}
        \centering
{\includegraphics[width=0.9\linewidth]{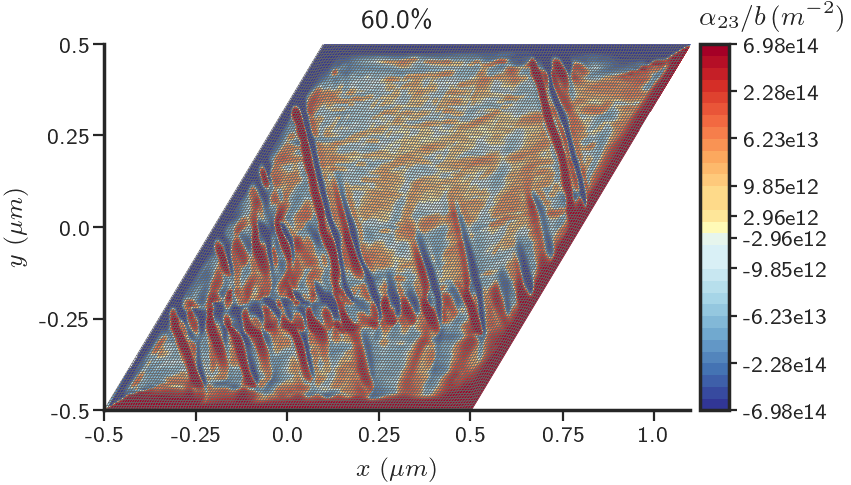}}
\caption{}
            \label{fig:p1_1mc_c_3slip_30_a23_mesh_fine}
    \end{subfigure}
       \caption{ Microstructure for the  $\mum{1}$ domain size at $60\%$ strain with plastically constrained boundaries and $\theta_0 = \mdeg{30}$ ($n_{sl} = 3$): a) $\rho_g$ b) $\alpha_{13}$ c) $\alpha_{23}$ computed with  the coarse mesh,  d) $\alpha_{23}$ computed with  the fine mesh.}
\label{fig:convergence_micro}
 \end{figure}

The norm of the dislocation density  $\rho_g$ and the components of $\bfalpha$ ($\alpha_{13}$ and $\alpha_{23}$) for the $\mum{1}$ domain size at $60\%$ strain for the coarse mesh are shown in  Figures  \ref{fig:p1_1mc_c_3slip_30_an_mesh}, \ref{fig:p1_1mc_c_3slip_30_a13_mesh}, and \ref{fig:p1_1mc_c_3slip_30_a23_mesh}, respectively. The localized concentrations are not aligned with the mesh. Moreover, the signed components are spread over more than $2$ elements in the mesh. Similar observation can also be made for the microstructure on the refined mesh, as shown in Fig.~\ref{fig:p1_1mc_c_3slip_30_a23_mesh_fine}. 

\begin{figure}[htbp]
    \centering
    \begin{subfigure}[b]{.495\linewidth}
        \centering
{\includegraphics[width=0.9\linewidth]{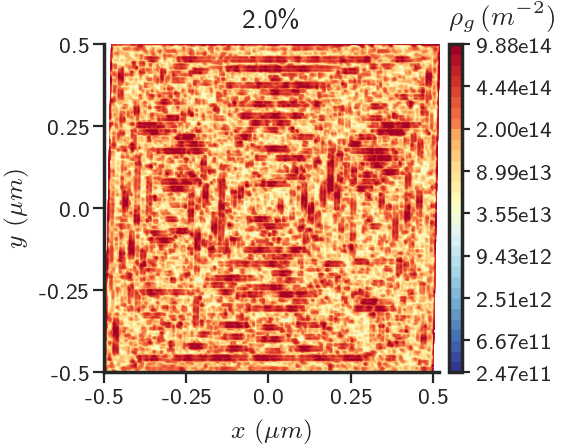}}
    \end{subfigure}%
    \begin{subfigure}[b]{.495\linewidth}
        \centering
{\includegraphics[width=0.9\linewidth]{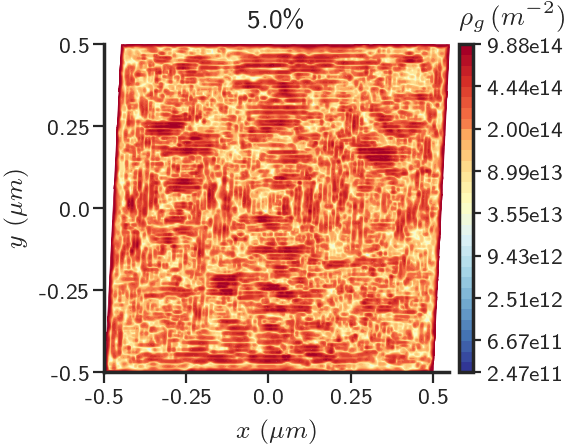}}
    \end{subfigure}\\
    \begin{subfigure}[b]{.495\linewidth}
        \centering
{\includegraphics[width=0.9\linewidth]{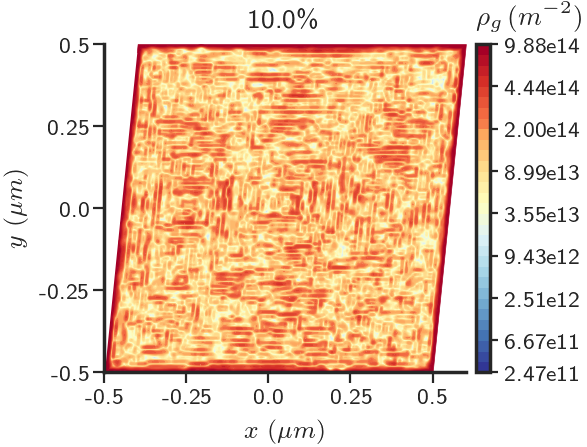}}
    \end{subfigure}%
    \begin{subfigure}[b]{.495\linewidth}
        \centering
{\includegraphics[width=0.9\linewidth]{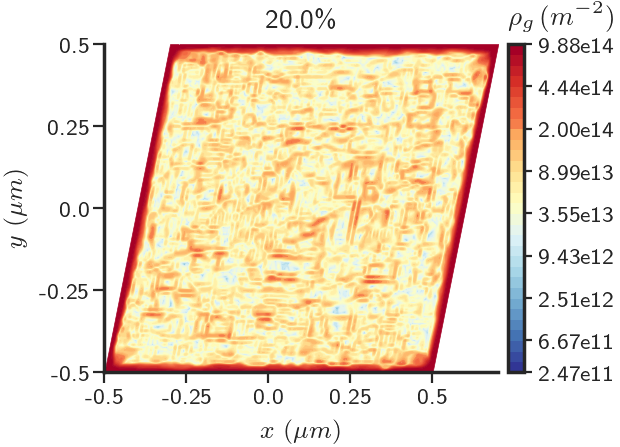}}
    \end{subfigure}\\
        \begin{subfigure}[b]{.495\linewidth}
        \centering
{\includegraphics[width=0.9\linewidth]{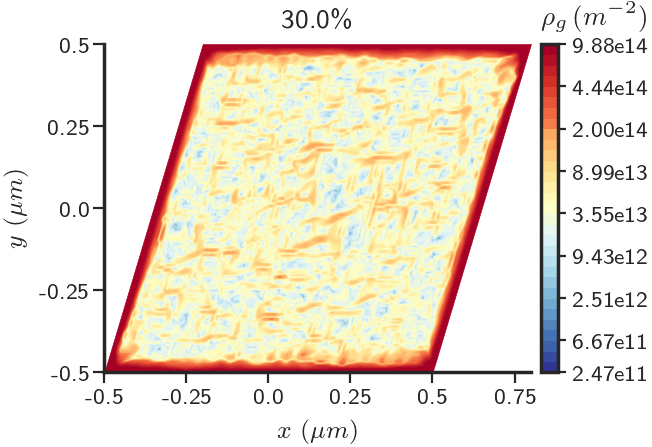}}
    \end{subfigure}%
    \begin{subfigure}[b]{.495\linewidth}
        \centering
{\includegraphics[width=0.9\linewidth]{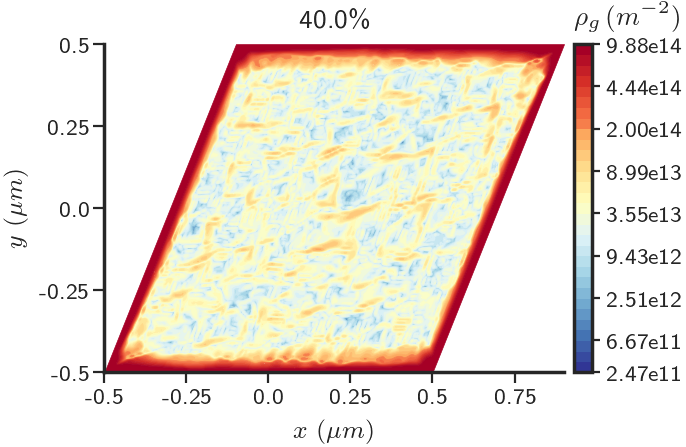}}
    \end{subfigure}\\
            \begin{subfigure}[b]{.495\linewidth}
        \centering
{\includegraphics[width=0.9\linewidth]{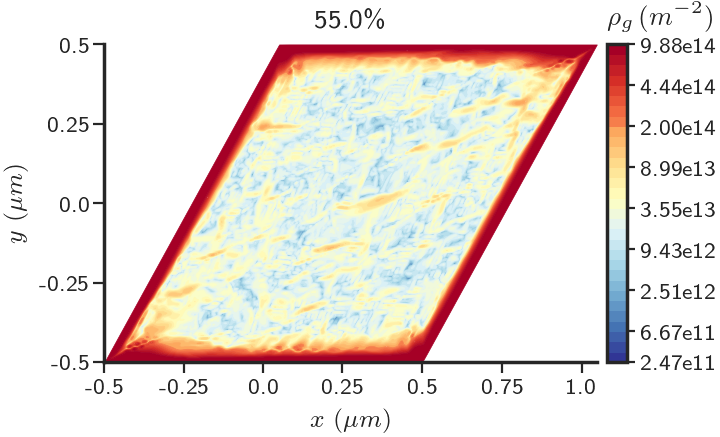}}
    \end{subfigure}%
    \begin{subfigure}[b]{.495\linewidth}
        \centering
{\includegraphics[width=0.9\linewidth]{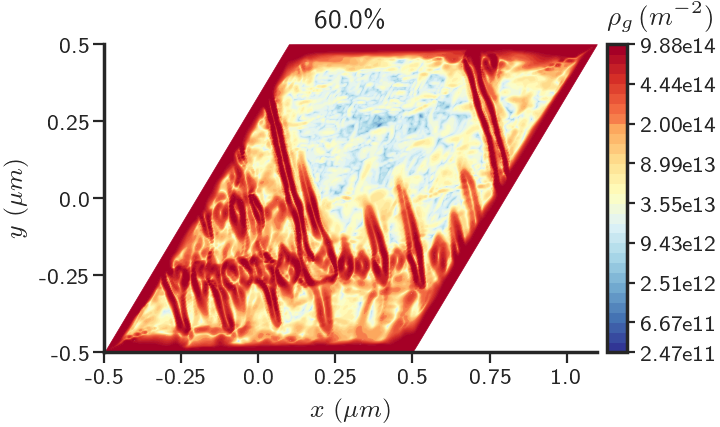}}
    \end{subfigure}%
    \caption{$\rho_g$  for the  $\mum{1}$ domain size at different strains with plastically constrained boundaries on the fine mesh with $\theta_0 = \mdeg{30}  ~(n_{sl} = 3)$.}
            \label{fig:p1_1mc_c_1slip_30_an_ref}
 \end{figure}

\begin{figure}[htbp]
    \centering
    \begin{subfigure}[b]{.495\linewidth}
        \centering
{\includegraphics[width=0.9\linewidth]{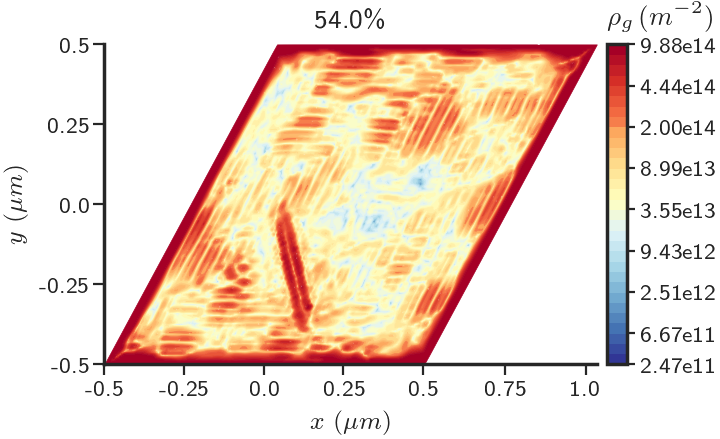}}
    \end{subfigure}%
        \begin{subfigure}[b]{.495\linewidth}
        \centering
{\includegraphics[width=0.9\linewidth]{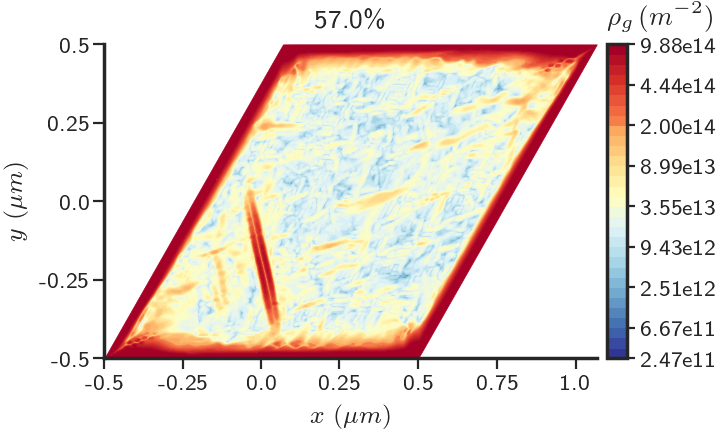}}
    \end{subfigure}\\
    \begin{subfigure}[b]{.495\linewidth}
        \centering
{\includegraphics[width=0.9\linewidth]{./figures/SE/3slipsystem/case1/C/1mic/AN-55pcnt.png}}
    \end{subfigure}%
    \begin{subfigure}[b]{.495\linewidth}
        \centering
{\includegraphics[width=0.9\linewidth]{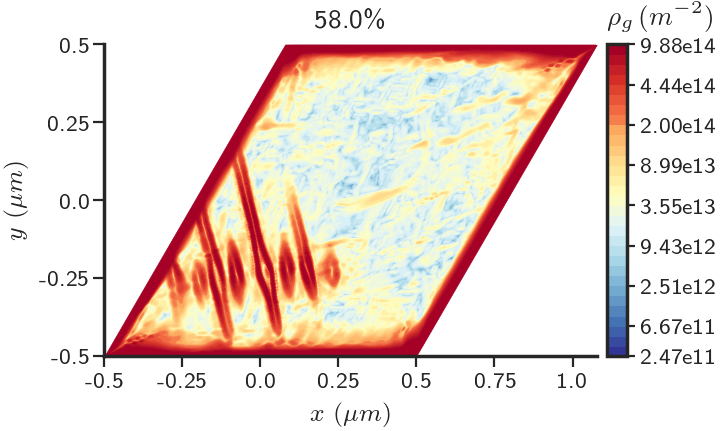}}
    \end{subfigure}\\
        \begin{subfigure}[b]{.495\linewidth}
        \centering
{\includegraphics[width=0.9\linewidth]{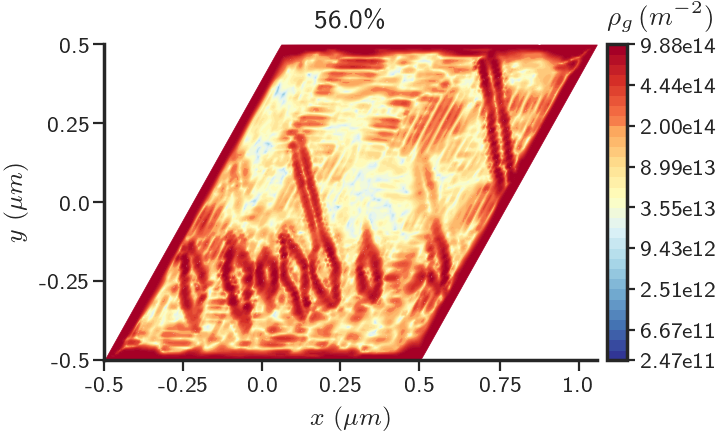}}
    \end{subfigure}%
    \begin{subfigure}[b]{.495\linewidth}
        \centering
{\includegraphics[width=0.9\linewidth]{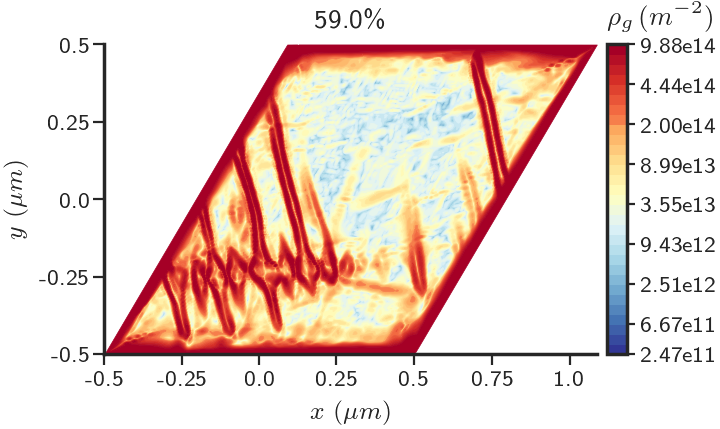}}
    \end{subfigure}\\
        \begin{subfigure}[b]{.495\linewidth}
        \centering
{\includegraphics[width=0.9\linewidth]{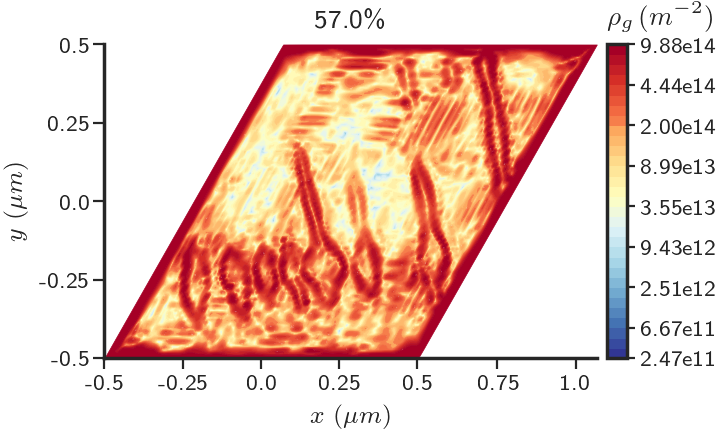}}
    \end{subfigure}%
    \begin{subfigure}[b]{.495\linewidth}
        \centering
{\includegraphics[width=0.9\linewidth]{./figures/SE/3slipsystem/case1/C/1refmic/AN-60pcnt.png}}
    \end{subfigure}%
    \caption{Results from the coarse mesh on left, and the fine mesh on right: Similarity of patterns for the  $\mum{1}$ domain size with plastically constrained boundaries and $\theta_0 = \mdeg{30}$ ($n_{sl} = 3$).}
\label{fig:self_similarity}
 \end{figure}

However, it is also clear, from a comparison of Figs.~\ref{fig:p1_1mc_c_3slip_30_an} and \ref{fig:p1_1mc_c_1slip_30_an_ref}, that even though the stress-strain curves converge for the coarse and the fine meshes, the microstructures are not converged for the mesh sizes considered. Nevertheless, we show that there are similarities in the microstructures obtained for the fine and coarse meshes considered at different levels of applied strain, as shown in Figure \ref{fig:self_similarity}.

\subsection{A necessary condition for microstructural patterns}\label{sec:k_0}
Our model predicts inhomogeneous distributions of dislocations leading to the formation of microstructural patterns such as dipolar dislocation walls and cell structures. The discussion surrounding \eqref{eq:LpXn_top} and \eqref{eq:LpXn_left} explains the reason for the generation of dislocation density in the constrained case, but the results of the unconstrained case in Sec. \ref{sec:unconstrained} begs the question of the real reason for the development of patterns in MFDM.

In \cite{roy2006size}, using the small deformation variant of MFDM, mild patterns were obtained for the $\mum{1}$ domain size with constrained and unconstrained boundaries. To understand the issue, a simplified system in 1 space dimension was analyzed in \cite[Sec.~4.2.1]{roy2006size} and linearized weak hyperbolicity of the homogeneous state was pointed out as a possible reason for giving rise to a (controlled) instability making the system sensitive to perturbations and leading to the formation of patterns. The homogeneous state may be thought of as a situation where the  $\bfalpha$ and $g$ evolution equations (\eqref{eq:mfdm_alpha} and \eqref{eq:softening} respectively) are uncoupled from each other instantaneously, which leads to the hypothesis that $k_0 = 0$ may lead to the suppression of patterns.

This hypothesis, i.e.~$k_0 = 0$ suppresses patterns, was tested  in \cite{das2016microstructure} using a simple $1$-$d$ ansatz, and it was again observed that the microstructure vanishes in the absence of any coupling between the dislocation transport \eqref{eq:mfdm_alpha} and the strength evolution equations \eqref{eq:softening}. 

\begin{figure}[htbp]
    \centering
    \begin{subfigure}[b]{.495\linewidth}
        \centering
{\includegraphics[width=0.9\linewidth]{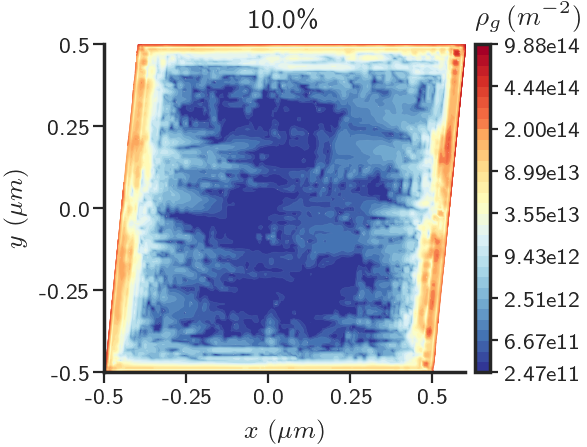}}
    \end{subfigure}%
    \begin{subfigure}[b]{.495\linewidth}
        \centering
{\includegraphics[width=0.9\linewidth]{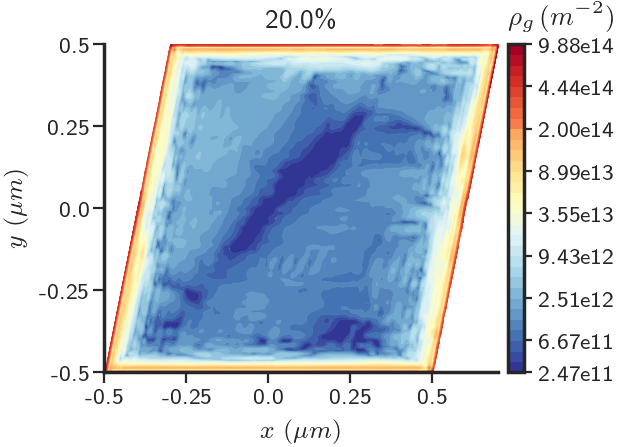}}
    \end{subfigure}\\
    \begin{subfigure}[b]{.495\linewidth}
        \centering
{\includegraphics[width=0.9\linewidth]{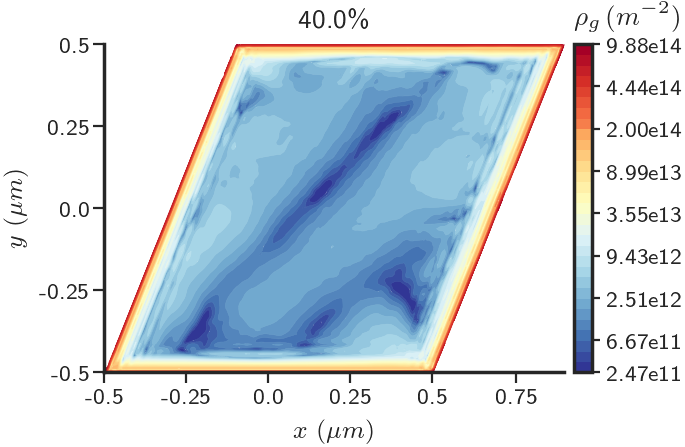}}
    \end{subfigure}%
    \begin{subfigure}[b]{.495\linewidth}
        \centering
{\includegraphics[width=0.9\linewidth]{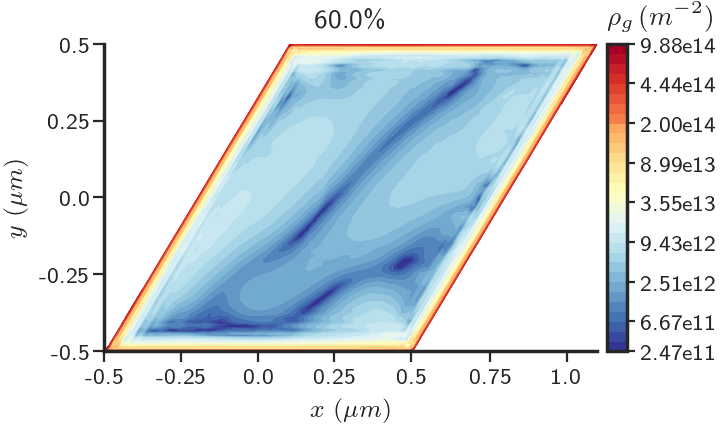}}
    \end{subfigure}%
        \caption{Distribution of $\rho_g$ for the $\mum{1}$ domain size at different strains with $k_0 = 0$ and plastically constrained boundaries $(\theta_0 = \mdeg{30},\,  n_{sl} = 3)$.}
            \label{fig:p1_1mc_c_3slip_30_an_k00}
 \end{figure}

Here, we test the same hypothesis in the finite deformation setting. $k_0 = 0$ is assumed, with all other parameters taken from Table \ref{tab:se_parameters} along with $\theta_0 = \mdeg{30}$ and $n_{sl} = 3$. Figure \ref{fig:p1_1mc_c_3slip_30_an_k00} shows the distribution of $\rho_g$ at different strains under such a scenario.  Comparing Fig.~\ref{fig:p1_1mc_c_3slip_30_an_k00} with Fig.~\ref{fig:p1_1mc_c_3slip_30_an}, we notice that the dislocation patterns entirely change when $k_0 = 0$; they are mildly patterned with much of the dislocations accumulated near the boundary similar to the case of the $\mum{5}$ sample size shown in Fig.~\ref{fig:p1_5mc_c_3slip_30_an_40pcnt}. Therefore, we conclude that a necessary condition for patterning in full finite deformation MFDM is the coupling between equations of dislocation transport and evolution of strength evolution through $k_0 \neq 0$.

\subsection{Effect of the length scale, \texorpdfstring{$l$}{l}}
\label{sec:role_of_l}
Here, we look at the effect of $l$, defined in \eqref{eq:Lp}, on the microstructure obtained during simple shearing of the $\mum{1}$ sample size with constrained boundaries, $\theta_0 = \mdeg{30}$, and $n_{sl} = 3$, with all other parameters as in Table \ref{tab:se_parameters}.

\begin{figure}[htbp]
        \centering
{\includegraphics[width=0.7\linewidth]{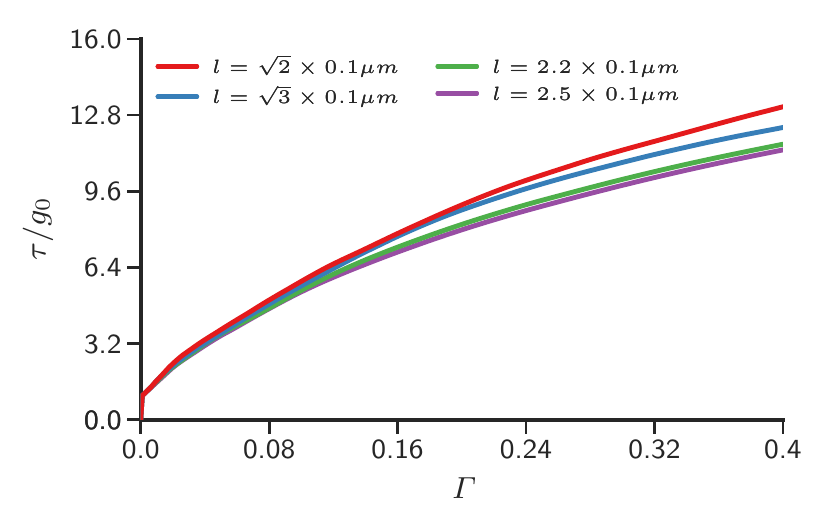}}
    \caption{Stress strain response for different values of $l$ for the  $\mum{1}$ domain size with plastically constrained boundaries ($\theta_0 = \mdeg{30},\,n_{sl} = 3$).}
    \label{fig:ss_leffect}
    \end{figure}

\begin{figure}[htbp]
    \centering
        \begin{subfigure}[b]{.495\linewidth}
        \centering
{\includegraphics[width=0.9\linewidth]{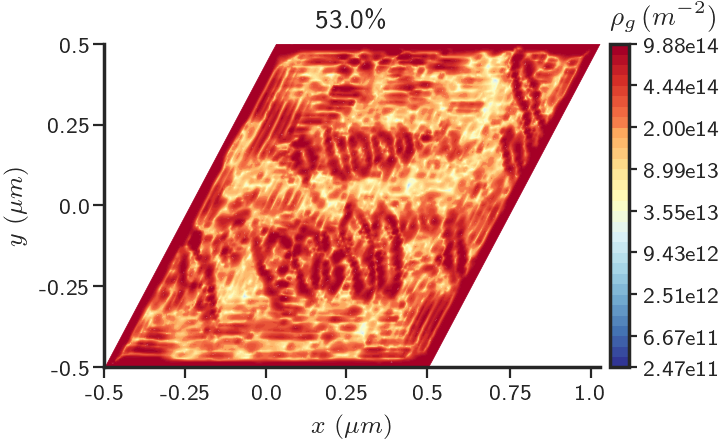}}
\caption{}
    \end{subfigure}%
    \begin{subfigure}[b]{.495\linewidth}
        \centering
{\includegraphics[width=0.9\linewidth]{./figures/SE/3slipsystem/case1/C/1mic/AN-60pcnt.png}}
\caption{}
    \end{subfigure}\\
    \begin{subfigure}[b]{.495\linewidth}
        \centering
{\includegraphics[width=0.9\linewidth]{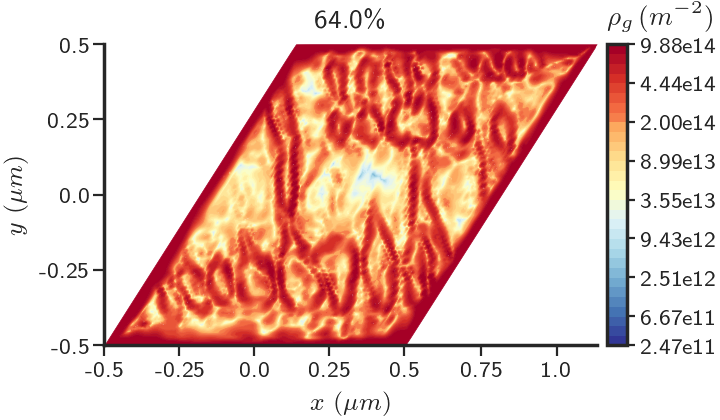}}
\caption{}
    \end{subfigure}
    \begin{subfigure}[b]{.495\linewidth}
        \centering
{\includegraphics[width=0.9\linewidth]{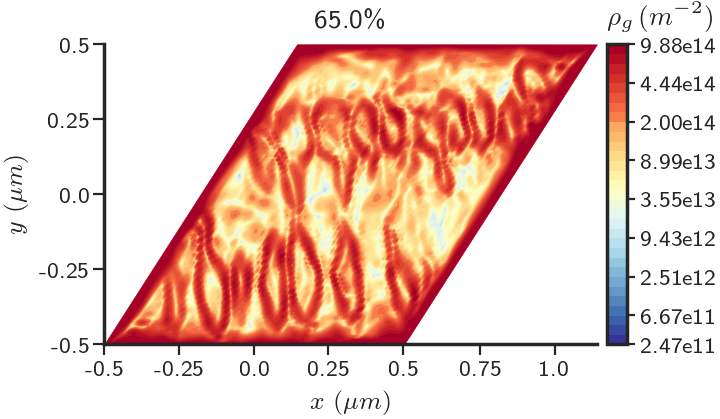}}
\caption{}
    \end{subfigure}%
        \caption{Distribution of $\rho_g$ for the $\mum{1}$ domain size with plastically constrained boundaries ($\theta_0 = \mdeg{30},\,n_{sl} = 3$) a) $l = \sqrt{2} \times 0.1 \,\mu m$ b) $l = \sqrt{3} \times 0.1 \,\mu m$ c) $l = 2.2 \times 0.1\, \mu m$ d) $l = 2.5 \times 0.1 \,\mu m$.}
            \label{fig:p1_1mc_c_3slip_30_an_l_effect}
 \end{figure}

We first look at the variation in stress-strain response for different values of $l$ shown in Fig.~\ref{fig:ss_leffect}. A decrease in the value of $l$ results in stronger response. As already explained, this is due to the fact that a larger $l$ decreases the magnitude of $\bfalpha$ in the domain and consequently leads to smaller hardening \eqref{eq:softening}. Next we look at the effect of $l$ on the microstructural patterns shown in  Figure \ref{fig:p1_1mc_c_3slip_30_an_l_effect}. It can be seen that increasing the value of $l$ makes a noticeable difference in the applied strain where qualitatively similar patterns of dislocations are formed. We notice that for $l = 2.5\times 0.1 \mu m$ we get several dislocation cells in the domain at $65 \%$ strain. Similar and even more intense structures can be noticed for $l = \sqrt{2} \times 0.1 \mu m$ at $53\%$ strains. Therefore, we can conclude that similar microstructures form at comparatively smaller strains as $l$ is decreased, and the distribution has higher magnitude on average as well.

\section{Concluding remarks}\label{sec:conclusion}
We have presented a first model of mesoscale crystal plasticity of unrestricted geometric and material nonlinearity in the literature, and used a finite element implementation of it to demonstrate dislocation patterning as well as size effects. The implementation is quite efficient, and a typical 2-d simulation up to $60 \%$ strain on the meshes shown in Table \ref{tab:se_parameters}  takes an average wall-clock time of $5$ hours when running on $1$ node comprising $24$ processors. Interesting and realistic microstructural features of plastic response have been shown to be within the qualitative purview of the model, which may be considered a minimal enhancement of classical crystal plasticity to account for what are commonly known as geometrically necessary dislocations. The general ideas involved in the development of the mesoscale model lend themselves to more refined descriptions, obviously with concomitant added cost.

While in this paper we have focused on dislocation microstructures that are decoupled from deformation microstructures, it is not our intent to downplay the importance of the latter. Our future work with this model will focus on dislocation patterning accompanying deformation microstructures like shear bands \cite{asaro1977strain, peirce1983material, peirce1983shear, ortiz1999nonconvex, aubry2003mechanics} and patchy slip \cite{piercy1955study, cahn1951} arising from the effects of strong latent hardening \cite{asaro1983micromechanics, bassani1993plastic}. Comparison with experiment of the evolving cellular and wall patterns formed will also be the subject of future work.

With this work, we hope to have moved the subject of plasticity to within the realm of nonlinear, pattern-forming, continuous-in-time dynamical systems without any non-standard restrictions like rate-independence or having to pose the problem in a time-discrete manner with an invocation of the direct methods of the calculus of variations within time-steps. A comprehensive large-scale computational study of the nature of convergence of the observed patterns in our work awaits further study, including whether weaker notions of convergence \cite{fjordholm2017construction, fjordholm2017statistical} will be required. Also, the significance and utility of the work of the French school of Mathematical Morphology \cite[e.g.,][]{jeulin2013analysis, angulo2017convolution} in understanding and characterizing the intricate patterns displayed by our model appears to be an interesting area of future research.

\subsubsection*{Acknowledgments}
This research was funded by the Army Research Office grant number ARO-W911NF-15-1-0239. The developed computational framework uses the following open source libraries:  Deal.ii \cite{dealII85}, P4est \cite{BursteddeWilcoxGhattas11}, MUMPS \cite{MUMPS:1}, and PetSc \cite{petsc-web-page}. This work also used the Extreme Science and Engineering Discovery Environment (XSEDE) \cite{xsede1}, which is supported by National Science Foundation grant number ACI-1548562. We gratefully acknowledge the Pittsburgh Supercomputing Center \cite{psc}, and Jorge Vin\~als and the Minnesota Supercomputing Institute (URL: http://www.msi.umn.edu) for providing computing resources that contributed to the research results reported within this paper. 



\clearpage
\newpage

\bibliographystyle{alpha}
\bibliography{gen_bib}

\end{document}